# Uninterrupted Maximum Flow on Signalized Traffic Networks

Melvin H. Friedman,[1] Brian L. Mark,[1] and Nathan H. Gartner[2]

## Abstract

This paper describes a traffic signal control strategy that allows motorists who travel at a recommended speed on suburban arterial two-way roads with a common cycle time to make every traffic signal. A road-to-traveler-feedback-device (RTFD) advises motorists how fast they should travel to do this. Signalized arterial roads where vehicles that travel at the recommended speed make every traffic signal are termed Ride-the-Green-Wave (RGW) roads. Left-turn-arounds enable vehicles to turn left from two-way RGW-roads to intersecting/orthogonal two-way RGW-road while allowing maximum flow at the intersection. The traffic signal control technique that enables vehicles that travel at the recommended speed to make every traffic signal has been verified using a simulation program (RGW-SIM). In addition to introducing novel traffic signal control strategies, the methods presented in this paper have implications for road network design, public transport control, connected and automated vehicles and environmental impacts.

## 1. Introduction

The predominant flow of traffic in urban street networks is along arterial roads and is constrained by traffic signals at intersections. Coordination of traffic lights along the arterials provides numerous advantages as indicated in the traffic engineering literature [1]. Two common approaches to coordinating arterial traffic signals are: 1) maximize progression bandwidth or 2) minimize overall delays and stops [2, 3].

The first mathematical optimization model for arterial bandwidth maximization used a combinatorial search scheme [4]. This model calculated offsets for two-way symmetric progression bands. Subsequently, a more advanced model was formulated using mixed-integer linear programming techniques [5]. This model optimizes progression speed and cycle time, in addition to offsets culminating in the MAXBAND model [6]. Advances in optimization techniques and computational capabilities have steadily increased the sophistication and versatility of traffic signal bandwidth optimization models [7-9]. Similar approaches have been introduced for the combined control of vehicular traffic and transit vehicles [10, 11]. Examples of delay-and-stops-based models can be found in a review paper [2]. The introduction of connected and automated vehicle (CAV) technologies offers new opportunities for traffic signal coordination [12]. Another approach to arterial progression optimization simultaneously takes into account through and left-turn traffic needs [13]. Tajalli and Hajbabaie [14] describe how to coordinate traffic signals with a mixture of connected automated vehicles and human-driven vehicles. Numerous other authors have discussed green wave methods, e.g., [15, 16].

All these models attempt to adjust the coordination progression to the prevailing traffic speed. Since such speeds are often highly variable, the resulting performance is often sub-optimal and network capacity is underutilized. The objective of this paper is to show how selected existing arterial roads can be converted into a network of roads that have the following properties: 1) vehicle platoons which travel at recommended speeds make all traffic signals, and 2) the arterial road network control strategies enables maximum flow. We refer to such roads as Ride-the-Green-Wave (RGW) roads. A patent application has been submitted to the US Patent Office describing an implementation for some of the material presented in this paper [17]. A one-page summary of this invention is available [18].

Concepts discussed in this paper that result in RGW roads in the context of our proposed green-wave approach to traffic signal coordination include: 1) left-turn-arounds, 2) road-to-traveler-feedback-device (RTFD), 3) RGW-traffic signals, 4) virtual nodes, 5) green-waves, 6) green-arrows, 7) left-turn-arrows, 8) reduced-capacity-arrows, 9) blue traffic signals, and 10) virtual nodes. On one- or two-way RGW-roads motorists, who travel at the recommended speed make all traffic signals regardless of the distance traveled (video-1). RGW-roads also have the maximum possible bandwidth. Bandwidth is the fraction of the common cycle time during which vehicles can travel uninterrupted through a series of signals. When maximum bandwidth is not needed, RGW-roads can trade bandwidth to gain more convenient left-turns. The left-turn-around concept allows vehicles to make left-turns while simultaneously allowing maximum vehicle flow in four orthogonal directions. The RTFD recommends motorist speeds, essential for optimum RGW-road performance, using the uninterrupted maximum flow methodology of this paper. GLOSA [19 - 21], Connected Automated Vehicles (CAV) [12, 14] and Transit Signal Priority [22] are alternative approaches for helping motorists travel through traffic signals with minimum interruptions.

There are two types of RGW-traffic signals: 1) real traffic signals located at intersections and 2) virtual traffic signals that are not at intersections. Virtual nodes are the locations on RGW-roads where virtual traffic signals are placed. The location of RGW-traffic signals (real or virtual) determines green-wave speed in conjunction with the traffic signal cycle time. One can think of the RGW-traffic signals as producers/generators of green-waves that travel at a speed determined by the distance between adjacent nodes and the cycle time. The basic idea is to set up green-waves that never stop moving and never intersect, which fully utilize each intersection. Instead of signals controlling traffic we think of traffic controlling the signals. Thus, conceptually, the motion of green-waves (which consists of vehicle platoons) turns traffic signals green when they enter an intersection and red when they leave it. In this paper green waves are represented by green arrows which travel at the same speed as green-waves but have the additional property of being able to turn traffic signals yellow or all-red. Virtual nodes serve to: 1) reduce green-wave speed when it is excessive, 2) reduce variations in green-wave speed and 3) coordinate traffic signals on RGW-road networks. Left-turn-arrows and reduced-capacity arrows are generalizations of green-arrows and work together to allow motorists to make a left-turn without having to stop for a traffic signal. Green-arrows, left-turn arrows and reduced-capacity arrows all move like green-waves but differ in how their motion changes traffic signals. Left-turn- and reduced-capacity-arrows are used to make left-turns more convenient when there is anisotropic vehicle flow demand. For motorists without an RTFD, blue traffic signals inform motorists who want to make a left-turn whether they can do so directly or need to use a left-turn-around. Left-turn-arounds are needed when there is a high traffic flow demand in the opposite direction. Compared to other uninterruptable road systems, such as freeways or expressways, RGW-roads have greater accessibility, and lower capital cost/km.

An essential requirement to coordinate traffic signals on arterial roads or road-networks is that they operate on the same cycle time. Uninterrupted maximum flow methodology requires/assumes: 1) the default state for all traffic signals on the RGW-network is red, 2) green-wave motion controls traffic signals: when a green-wave enters an intersection it turns traffic signals green and as it leaves the intersection it turns the signal amber and then red, 3) green-waves have the following properties: A) Green-waves never stop moving. B) Green-waves fully utilize each intersection. C) Green-waves never intersect one another. Logical consequences of requirements A) – C) imply Results 2.x and Results 3.y stated in Sections 2 and 3, respectively. A Mathematica program, RGW-TLP (Ride the Green Wave Traffic Light Progression), takes as input cycle time and node locations to produce green-wave speed, green time in the forward direction $T_{gf}$ and reduced-offset-times $T_{roffset}$ that determine traffic signal timing [23]. Finding optimum RGW-node locations can be done by minimizing a function $\eta$ subject to constraints as described in Section 4. Here, cycle time and RGW-node locations were chosen with the objective of minimizing $\eta$ and are applied (as an example) to Telegraph Road located in Alexandria, VA, USA in Section 4. The Uninterrupted Maximum Flow procedure illustrated here for a single two-way road has been further generalized to a network of two-way suburban roads [23].

The methodology developed in this paper has been validated by a simulation program RGW-SIM2 (Section 5 and Appendix A) which was also used to estimate how well RGW-roads work for real drivers and CAV.

## 2. Cartesian grid two-way roads

We introduce first a Cartesian grid of two-way roads. These results will then be extended to more general road networks.
RGW-roads, RGW-nodes, green-waves, and green-arrow concepts described in Section 1 are essential to understanding this section. Green-waves, here represented by *green-arrows*, are conceptual entities that travel on arterial roads with the property that when they enter an intersection they turn red-lights green and as they leave the intersection the tail of the arrow turns the light amber and then red. In this work *green-arrows* represent potential vehicle platoons. Green-arrows can be: a) empty or nearly empty when traffic is light or b) they can have a density given approximately by Figure 1 under stable-saturation-flow, or c) they can be oversaturated when vehicle density is substantially greater than that given by Figure 1. Green-arrow laws-of-motion and driver behaviour imply that under conditions a) and b) motorists can travel the entire length of RGW-roads, regardless of length, and make every traffic signal without stopping.

*Green-arrow laws-of-motion*. Five green-arrow laws-of-motion on a Cartesian two-way road-network are postulated. 1) Green-arrows never stop moving. 2) On two-way arterial roads, whenever a green-arrow enters an intersection another green-arrow traveling in the opposite direction enters the same intersection. 3) Whenever a green-arrow leaves an intersection another green-arrow traveling in an orthogonal direction enters the intersection, 4) Green-arrows never intersect one another. 5) Green-arrows on a regular Cartesian road-network travel at a constant speed. The first law ensures green-arrows have uninterrupted/continuous flow. The second and third laws ensure each intersection is fully utilized. The fourth law is a fundamental traffic safety requirement since intersecting green-arrows would imply that vehicles in the waves represented by these arrows would collide with one another. The fifth law, which will later be modified to accommodate existing roads, is imposed for simplicity and symmetry: 1) on a Cartesian grid speed in the north-south direction is assumed to be the same as the speed in the east-west direction, and 2) in an uninterrupted flow system, where vehicles never stop, speed is assumed to be constant. A virtue of the fifth law is that it is readily extended to roads not described by a Cartesian grid.

We invoke two common drivers' rules-of-behaviour. 1) *Allow one car length between trailing and leading vehicle for every 16 kph of vehicle speed*. 2) *The front bumper of a motorist's car should pass a fixed object two seconds after the rear bumper of the car ahead passed the same object* [24]. These rules of behaviour as they relate to traffic flow theory are confirmed by surveys [25].

*Result 2.1*. Using either driver rule-of-behaviour: a) more vehicles can fit in a green-arrow as vehicle speed decreases, b) safe vehicle speed decreases as vehicle road density increases, c) when road density is high, by suitably adjusting vehicle speed so as to satisfy either rule-of-behaviour, vehicle flow rate [veh hr$^{-1}$ ln$^{-1}$] remains nearly constant. This is called the *stable-saturation-flow-rate*. Assertion c) is approximately valid because stable-saturation-flow-rate is proportional to the product of vehicle density and speed and as vehicle density goes up, vehicle speed goes down. Results 2.1 a) and b) are confirmed using published data [26] which is shown on a single graph in Figure 1 to emphasize how well Lincoln tunnel and Merritt parkway data sets agree with one another rather than on two separate graphs as done by Greenberg. Greenberg expressed his results in standard US units and we expressed his results in SI units. The model curve (1) was fit to the combined Merritt Parkway and Lincoln Tunnel data using a least squares fit with Mathematica to exhibit experimental data which support Result 2.1.

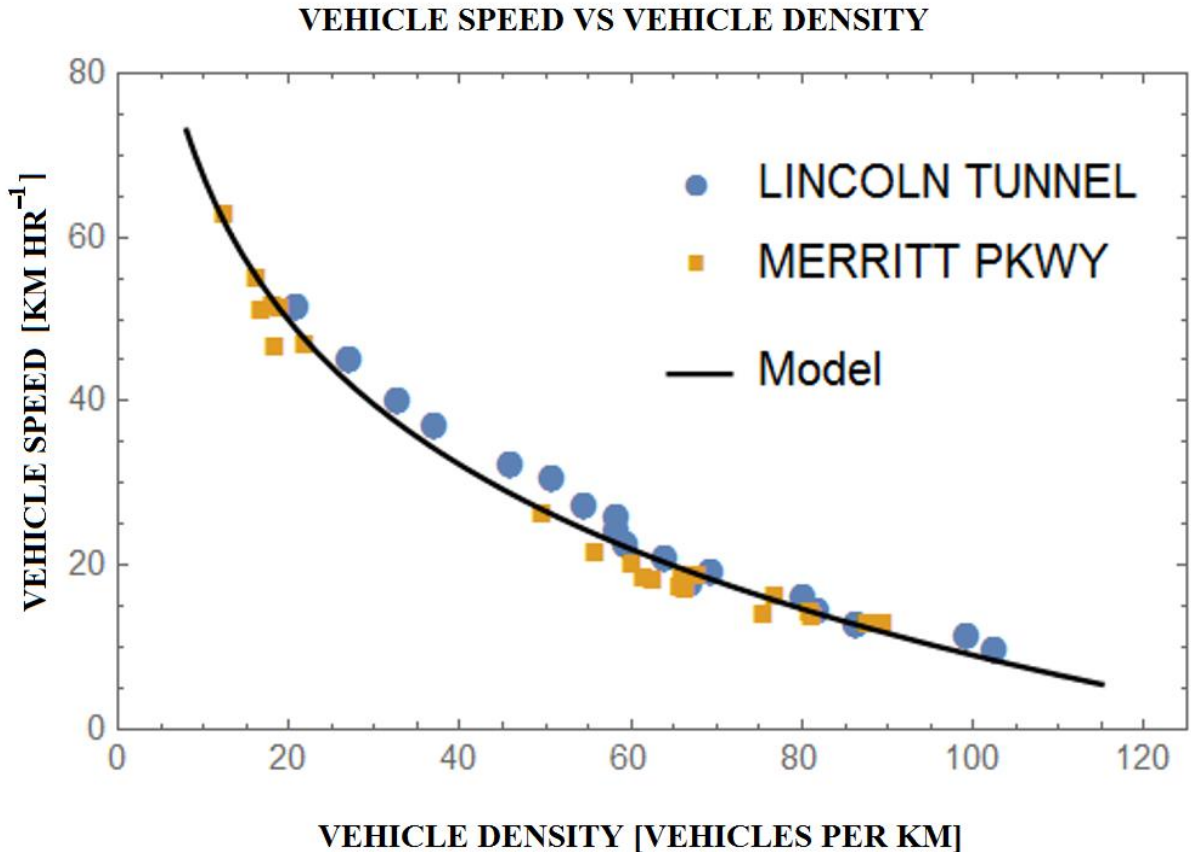

Figure 1. Experimental vehicle speed as a function of vehicle road density.

Results 2.1 a) and b) are confirmed in Figure 1, graphed here with Mathematica using Greenberg's published data [26]. The model curve (1) is a least squares fit to the combined Merritt parkway and Lincoln tunnel data using Mathematica's "Fit" function

$$u = 125.75 - 25.35 \ln \rho, \quad 8 < \rho < 115 \tag{1}$$

Here $u$ is vehicle speed [km hr$^{-1}$] and $\rho$ is vehicle density [veh km$^{-1}$]. Using the relationship $q = \rho u$, data exhibited in Figure 1 is used to create a graph of vehicle flow $q$ versus vehicle density (Figure 2).

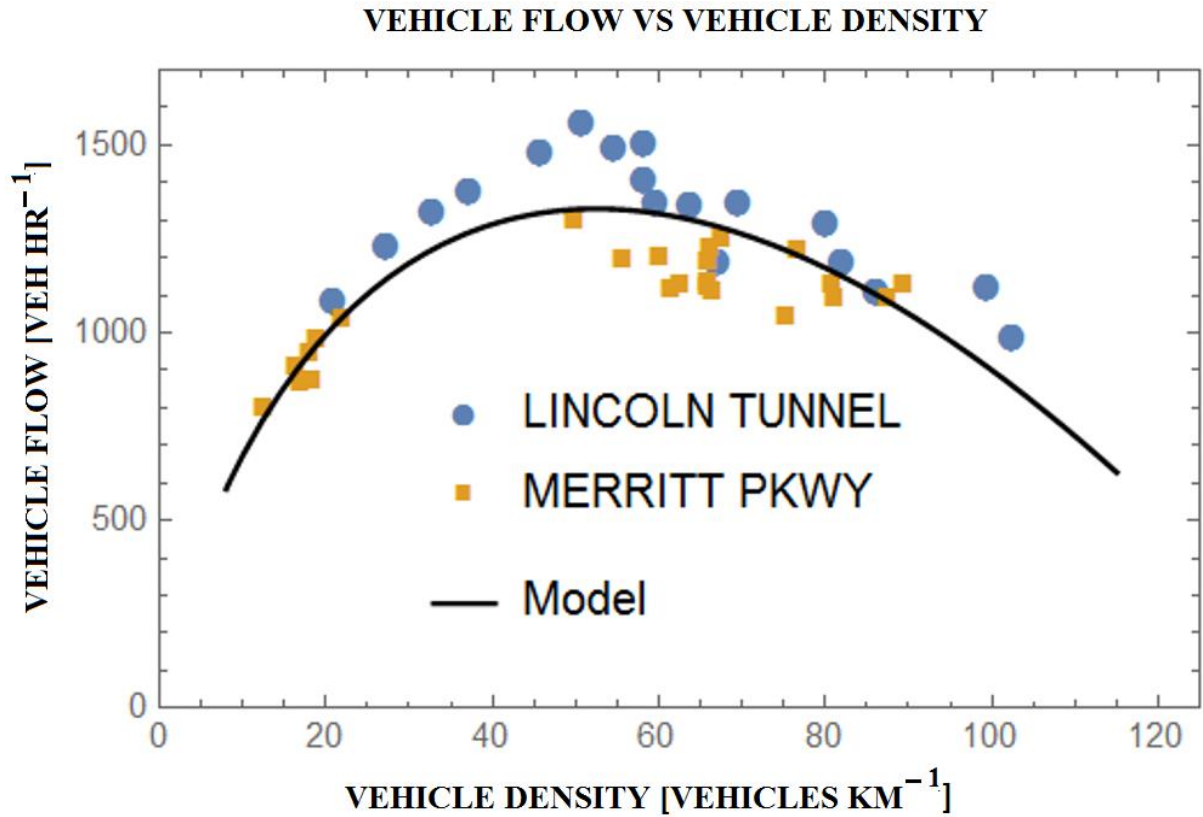

Figure 2. Vehicle flow as a function of vehicle density.

The model curve in Figure 2 is described by

$$q = \rho(125.75 - 25.35 \ln \rho), \quad 8 < \rho < 115) \tag{2}$$

The broad maximum in the data and model in Figure 2 for $\rho$ between 31 and 77 veh km$^{-1}$ supports result 2.1 c. The peak modelled flow $q = 1330$ veh hr$^{-1}$ ln$^{-1}$ takes place when $\rho$ is 52 veh km$^{-1}$. Other traffic flow models, similar to Greenberg's result can be found in the literature [27 - 30]. Work describing the fundamental diagram with mixture of human-driven and connected automated vehicles has been done by Zhou and Zhu [31] and by Yao et al. [32].

Figure 3 and video-1 [23, 33] demonstrate how green-waves with the above specified green-arrow laws-of-motion operate. The objective of our work is to coordinate traffic signals to achieve maximum uninterrupted stable-saturation-flow on suburban road networks. Because each intersection and each lane are utilized fully at each moment of time, this figure illustrates maximum flow; because green-waves/green-arrows never stop moving it illustrates uninterrupted flow; because the pattern is periodic, it applies to an arbitrarily large network. Figure 3 represents a top-down view of a city where green-waves are represented by green-arrows. Video-1/Figure 3 demonstrate that evenly spaced RGW-nodes implies vehicles travelling at the proper constant-speed make every traffic signal which reduces fuel consumption and pollution. The observation that evenly spaced traffic signals enables uninterrupted maximum flow has implications for city design and public transportation. (The assumption that RGW-nodes are evenly spaced is relaxed later in this paper.) Traffic signal state is determined by green-arrow motion. At the end of green-arrows, the shaft has amber and red portions. The shaft refers to the long, slender, cylindrical rod that forms the main body of a physical arrow. The moment a green-arrow enters an intersection the traffic signal at that intersection turns green and as the arrow leaves an intersection the amber/red portion of the shaft turn traffic signals amber/red. The time duration that the red-portion of a green-arrow is in an intersection is termed *all-red-time* in this paper and is also referred to as *red clearance time* in the literature. Figure 3 and video-1 were simplified by omitting the red/amber portions of green-arrows.

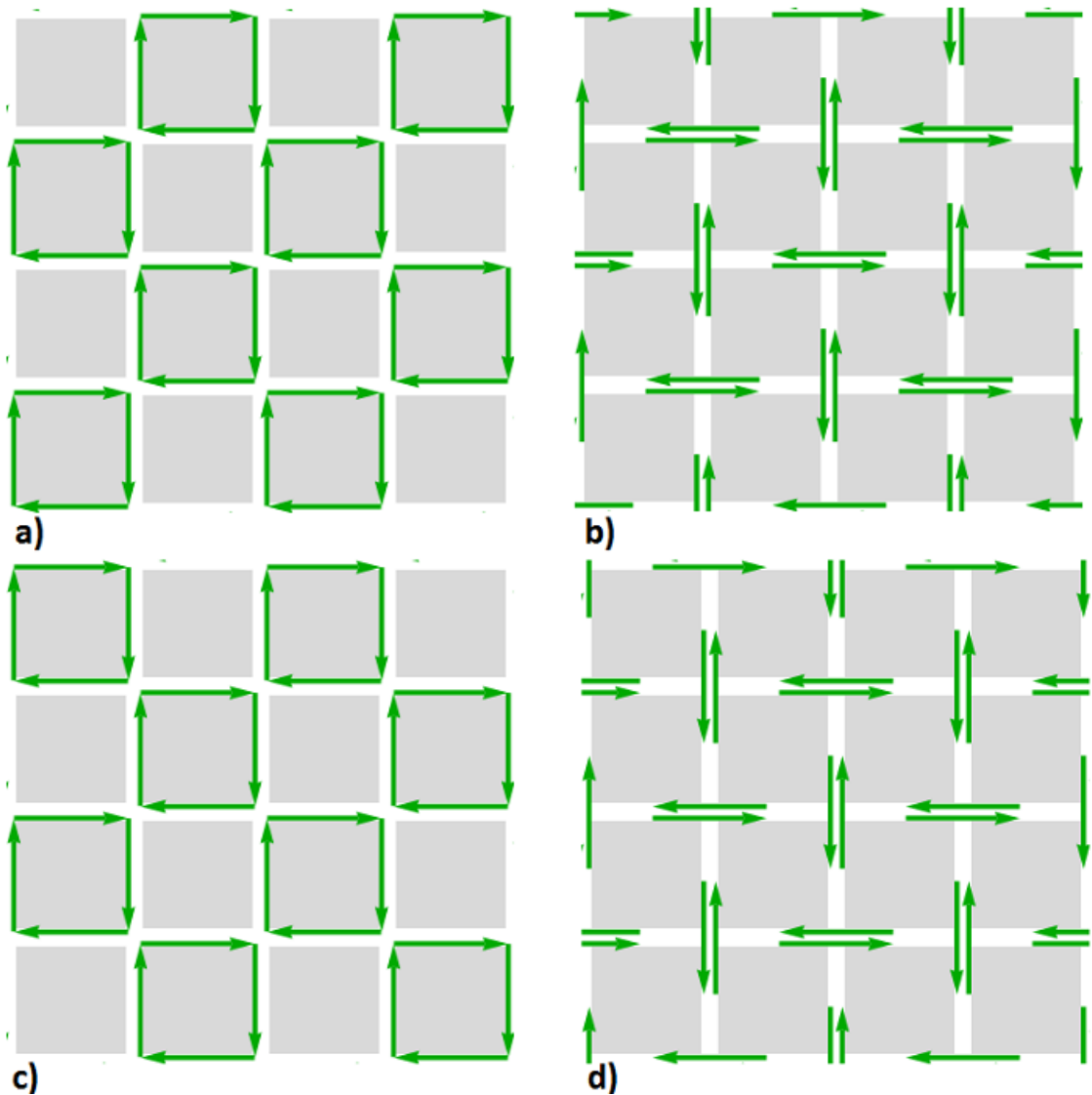

Figure 3. Four frames from video-1.

In Figure 3: a) the initial position of green-arrows; b) green-arrows advance ½ block from a); c) green-arrows advance one block from a); d) Green-arrows advance 1½ block from a). Advancement of arrows in d) by another ½ block produces a).

Grey-squares in Figure 3 are termed *blocks* with side length $L_b$. The white areas between grey-squares are termed *RGW roads*. Since green-waves never stop, vehicles which can maintain a position in a moving green-arrow make all traffic signals. Locations where RGW-roads intersect are termed *RGW-nodes*. Green-arrow length $L_g$ equals the distance between RGW-nodes; $L_g \cong L_b$ is a good approximation whenever road width is negligible compared to $L_b$. Figure 3 shows green-arrows have length one when measured in terms of block length. Since green-waves move at a constant speed, *green-time* $T_g$, the time a traffic signal is green, amber and all red, is proportional to $L_g$. *Red-time* $T_r$, the length of time traffic signals at RGW-nodes are red, corresponds to the white space between green-arrows. For isotropic flow $T_g = T_r$. The following results are deduced from Figure 3 and the green-arrow laws-of-motion.

***Result 2.2***. The location of a single green-arrow at an arbitrary instant of time determines the location of every green-arrow at that instant of time and for all subsequent time providing cycle time does not change.

***Result 2.3***. Green-wave speed is $v_g = L_g/T_g \cong L_b/T_g$. Since block length $L_b$ and green-time $T_g$ can be chosen subject to some constraints, green-wave speed is adjustable.

***Result 2.4***. For isotropic flow (flow independent of direction), cycle time $T_{cycle} = T_r + T_g = 2T_g$.

***Result 2.5***. Green-arrow lengths of one are the largest possible for isotropic flow on the Cartesian road-network exhibited in Figure 3. Green-arrow lengths greater than one inevitably violate green-arrow laws-of-motion 4.

***Result 2.6***. Green-arrow lengths (measured in terms of block length) are limited to discrete lengths of $1/n, n = 1, 2, 3, \ldots$ for two-way isotropic flow on the Cartesian road-network exhibited in Figure 3. Only for these lengths can green-arrow laws-of-motion be satisfied.

***Result 2.7***. Let *P* denote a midblock position on an RGW-road. For isotropic flow, illustrated in Figure 3, a green-wave is always moving past point *P* which is called a singular point. Thus, traffic signals, walkways and roads cannot be placed midblock.

***Result 2.8***. Figure 3 shows a finite area. However, Figure 3 can be extended to an arbitrarily large area or be reduced to a single RGW-node and still satisfy green-arrow laws-of-motion.

***Result 2.9***. Green-arrows traveling in two opposite directions enter RGW-nodes at the same moment.

In practical applications, to accommodate non-RGW-roads which cross RGW-roads, it is necessary to place a traffic signal *P* between RGW-nodes. For this reason, it is necessary to have equations that describe how long traffic signals are green $T_{gf}$ in the direction of RGW-roads and how long they are green $T_{gx}$ in the cross-direction to RGW-roads. The result depends on how far $\xi$ the traffic signal *P* is from an RGW-node.

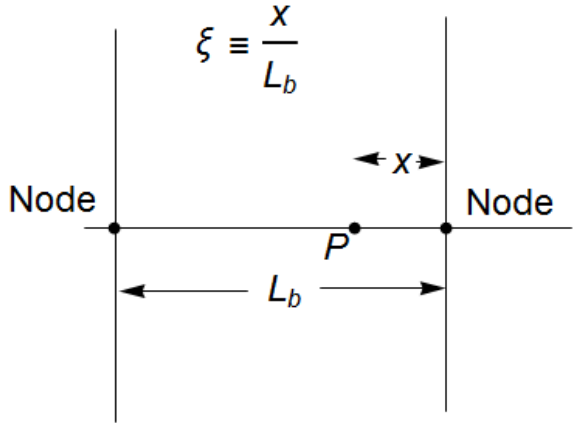

Figure 4. Definition of $\xi$.

In Figure 4, denote the distance of $P$ to the nearest RGW-node by $x$. The distance between RGW-nodes is green-arrow length $L_g$ but as discussed in Result 2.3 this is approximately equal to $L_b$. Then $\xi$, a dimensionless number, is defined as the ratio of $x/L_b$. From Figure 4 it is apparent that $\xi = 0$ corresponds to an RGW-node, $\xi = 1/2$ corresponds to the block-midpoint, and $0 \leq \xi < 1/2$. Figure 3 and green-arrow laws-of-motion on a Cartesian grid imply results 2.9 and 2.10 below.

**Result 2.10**. $T_{gf}(\xi) = (1 + 2\xi)T_g, \quad 0 \leq \xi < 1/2$

**Result 2.11**. $T_{gx}(\xi) = (1 - 2\xi)T_g, \quad 0 \leq \xi < 1/2$

Results 2.9 and 2.10 express *how long* traffic signals are green between RGW-nodes in the forward and cross directions. When $\xi = 0$ the green-forward-time $T_{gf}$ reduces to $T_g$ and when $\xi = 1/2$ the cross green-time $T_{gx}$ reduces to zero in agreement with Result 2.7. Observe that $T_{gf}(\xi) + T_{gx}(\xi) = 2T_g = T_{cycle}$, the traffic signal cycle time.

*Offset-time* $T_{offset}$ is the time a green-arrow takes to get to a specified intersection from a reference point and can be arbitrarily large. *Reduced-offset-time*

$$T_{roffset} \stackrel{\text{def}}{=} Mod[T_{offset}, T_{cycle}]$$

is defined as the time in a cycle when a green-arrow gets to a specified intersection from a reference point. By definition, reduced–offset-time is always between 0 and the cycle time.

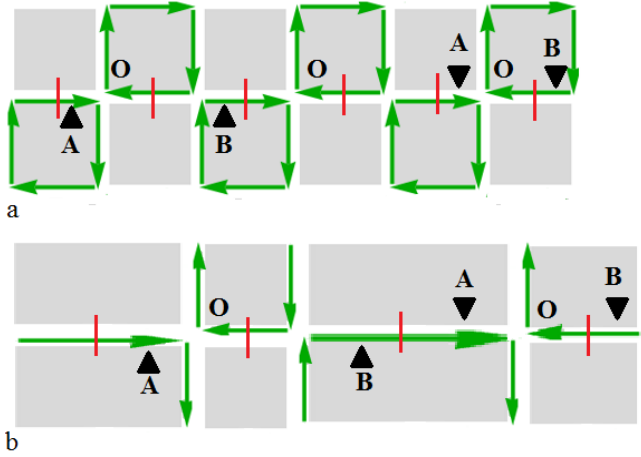

Figure 5. Illustration of *RGW-zero-offset-nodes* and *d parameter* for uniform and non-uniform Cartesian grids. Points A, B represent locations where RGW-roads intersect non-RGW-roads. Points O represent RGW-zero-offset-nodes.

Results 2.9 describes the magnitude of $T_{gf}$. The *d parameter* and *RGW-zero-offset-nodes*, described below and in Figure 5, are needed to describe *when* traffic signals turn green between RGW-nodes.

Figures 5 a) and b), respectively, show uniform and non-uniform Cartesian grids. Intersections marked with O, where green-arrows simultaneously enter an intersection, are, without loss of generality, termed *RGW-zero-offset-nodes*. Midblock locations are designated with red vertical line segments. The *d parameter* is the distance of a point (A or B in Figure 5) from the nearest RGW-zero-offset-node measured in units of blocks. It is a dimensionless number constrained by the relationship $0 < d < 1$. Points designated $A$ ($B$) denote $d$ parameters less (more) than ½. The extension to non-uniform non-Cartesian grids is immediate.

Figures 6 and 7 facilitate understanding equations that describe *when* traffic signals change for the case where traffic signals are between RGW-nodes. Traffic signals in Figure 3 turn red (green) periodically with a cycle time $T_{cycle}$. An *RGW-zero-offset-node,* at an instant of time, is arbitrarily defined when green-arrows traveling in opposite directions enter an intersection. In Figure 6 points labelled O are representative RGW-zero-offset-nodes. Examination of Figures 6 or 7 show that RGW-zero-offset-nodes are separated by two blocks. Let $d$ denote the nearest distance between RGW-node traffic signal $P$ to an RGW-zero-offset-node.

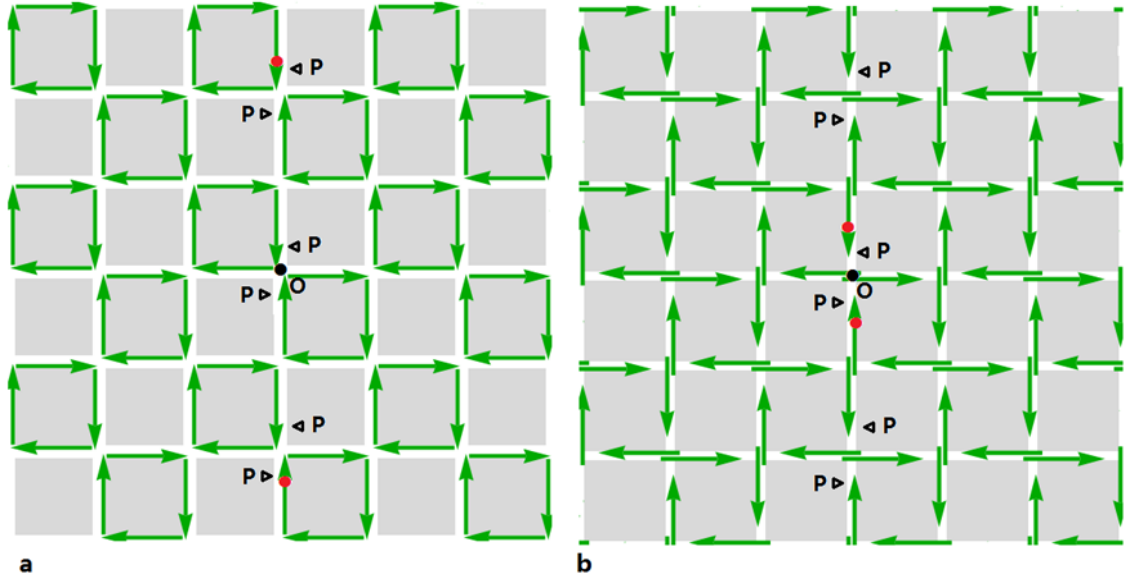

Figure 6. Calculation of $T_{roffset}$ for points $P$ (which denote traffic signal positions) north or south of a $T_{roffset} = 0$ RGW-node that satisfy the relationship $0 < d < 1/2$.

In Figure 6, a) denotes initial green-arrow position and b) denotes green-arrow position at a later time. In Figure 6 a) the RGW-node marked O has $T_{roffset} = 0$ in the north-south direction and points marked $P$ are ones where $T_{roffset}$ is calculated. Marker arrows in the northbound and southbound directions are designated with a red dot. Figure 6 b) shows the position of green-arrows at time given by Result 2.11.

***Result 2.12***. From Figure 6 it is apparent that for points $P$ *north or south* of an RGW-zero-offset-node the reduced-offset-time is $T_{roffset} = T_g(2 - \xi)$ when $0 < d < 1/2$.

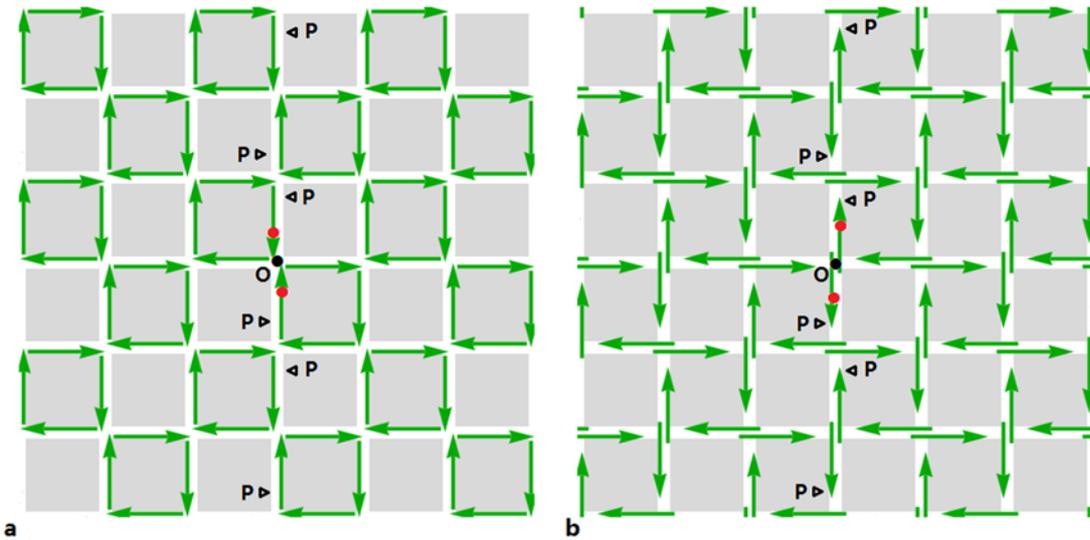

Figure 7. Calculation of $T_{roffset}$ for points $P$ (which denote traffic signal positions) north or south of $T_{roffset} = 0$ RGW-nodes that satisfy the relationship $1/2 < d < 1$.

Figure 7 a) denotes initial green-arrow position, and b) denotes green-arrow position at a later time. In Figure 7 a) the RGW-node marked O has $T_{roffset} = 0$ in the north-south direction and points marked P are ones where $T_{roffset}$ is calculated. Marker arrows in the northbound and southbound directions are designated with a red dot. Figure 7 b) shows the position of green-arrows at time given by Result 2.13 below.

**Result 2.13.** It is apparent from Figure 7 that for points P *north or south* of an RGW-zero-offset-node
$T_{roffset} = T_g(1 - \xi)$ when $1/2 < d < 1$.

The north-south and east-west symmetry in Figures 6 and 7 imply the validity of Results 2.13 and 2.14 below.

**Result 2.14.** For points P *east or west* of an RGW-zero-offset-node the reduced-offset-time $T_{roffset} = T_g(2 - \xi)$ when $0 < d < 1/2$.

**Result 2.15.** For points P *east or west* of an RGW-zero-offset-node the reduced-offset-time $T_{roffset} = T_g(1 - \xi)$ when $1/2 < d < 1$.

Results 2.11 – 2.14 are shown graphically in Figure 8.

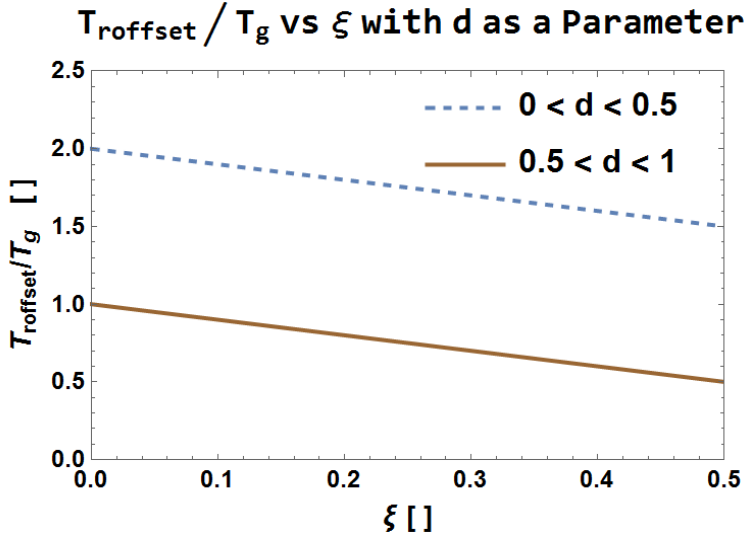

Figure 8. Graphical illustration of results 2.11 − 2.14.

Results 2.2 − 2.14 apply to *isotropic flow* where green-arrows in the north-, south-, east-, and west-directions are the same length. Now extend these results to *anisotropic flow* where green-arrow length is not the same in four orthogonal directions. Let $\alpha$ and $\beta$ denote green-arrow lengths in two orthogonal directions with $\alpha, \beta > 0$. One can satisfy green-arrow laws-of-motion with non-isotropic flow. For *non-isotropic flow* $\alpha \neq \beta$.

***Result 2.16***. Results 2.2, 2.3, 2.4, 2.8 and 2.9 are true for both isotropic and anisotropic flow. On two-way roads the largest value of $s \equiv \alpha + \beta = 2$.

***Result 2.17***. The green-arrow laws-of-motion motion can be satisfied over a continuum of $\alpha, \beta$ values subject to the constraints $\alpha, \beta > 0, \alpha + \beta = 2$.

***Result 2.18***. The value of $s \equiv \alpha + \beta$ is restricted to discrete values of $2/n, \; n = 1, 2, 3, \ldots$.

***Result 2.19***. The point $P$ in result 2.7 expands from a point to an interval on the road where $\alpha > \beta$.

***Result 2.20***. Green-wave speed $v_g = L_g/T_g$, but $L_g$ and $T_g$ are different in orthogonal directions.

The significance of anisotropic flow results will now be discussed. *Result 2.19* implies it will be difficult on existing arterial roads to obtain increased flow by simply increasing green-arrow length in one direction and decreasing it in the orthogonal direction as demanded by Result 2.17. Additional research on isotropic and anisotropic flow is in progress [33].

In this paper, when Figure 3 is applied to existing roads, because of Result 2.19, our analysis is confined to the case $\alpha = \beta = 1$. To coordinate traffic signals on existing roads, results 2.2, 2.3, 2.6, 2.8, 2.9, 2.10, 2.11, 2.12, 2.13 and 2.14 are explicitly needed.

A virtue of the methodology described so far is that when all green-arrows hold as many vehicles as they can (Figure 1), uninterrupted maximum vehicle flow is obtained. The virtue was obtained at a cost: 1) no provision has been made to include left-turns and 2) the methodology is only applicable to Cartesian road-networks. The first cost is addressed now; the second cost is handled in the next section.

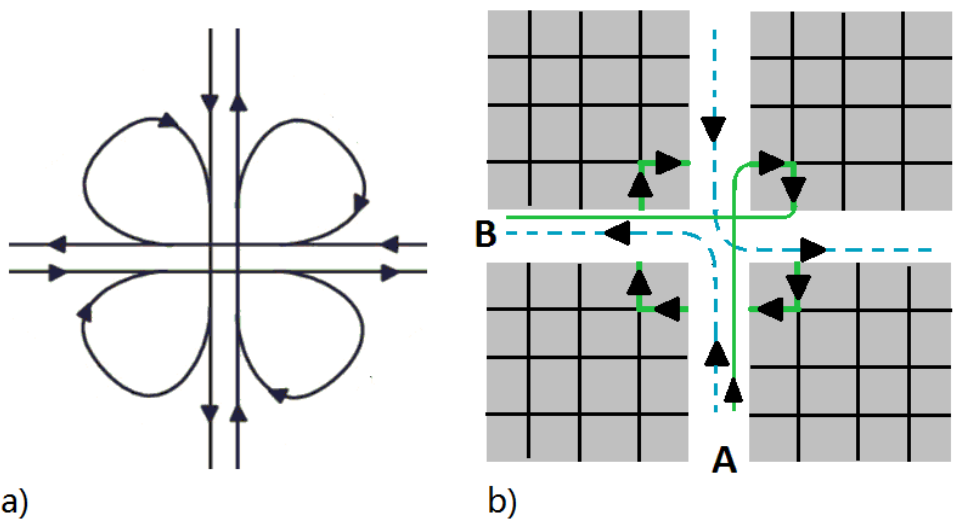

Figure 9. Direct- and cloverleaf-left-turns.

Figure 9 a) illustrates a cloverleaf-left-turns on a highway and Figure 9 b) illustrates direct- and cloverleaf- left-turns on RGW-roads. Black lines within blocks correspond to side-streets. Dashed and solid lines in Figure 9 b), respectively, illustrate direct- and cloverleaf-left-turns. The left-turn-around (LTA) consists of streets/roads where cloverleaf-left-turns are made. In urban areas existing streets provide the infrastructure for left-turn-arounds but in suburban areas left-turn-arounds may requires new infrastructure.

Examination of Figure 3 or video-1 [23, 33] shows that because green-arrows are always going through each intersection (either in the north-south or east-west directions), neglecting all-red time, maximum flow is obtained all the time for each intersection. Thus, direct-left-turns imply reduced flow. With the cloverleaf-left-turn, vehicles use existing streets/roads to make left-turns without reducing flow.

When traffic is light, direct-left-turns are more convenient but at times where maximum flow is needed cloverleaf-left-turns are needed. Figure 10 shows how motorists know: 1) they are on an RGW-road, 2) where left-turn-arounds are and 3) when they can make direct-left-turns.

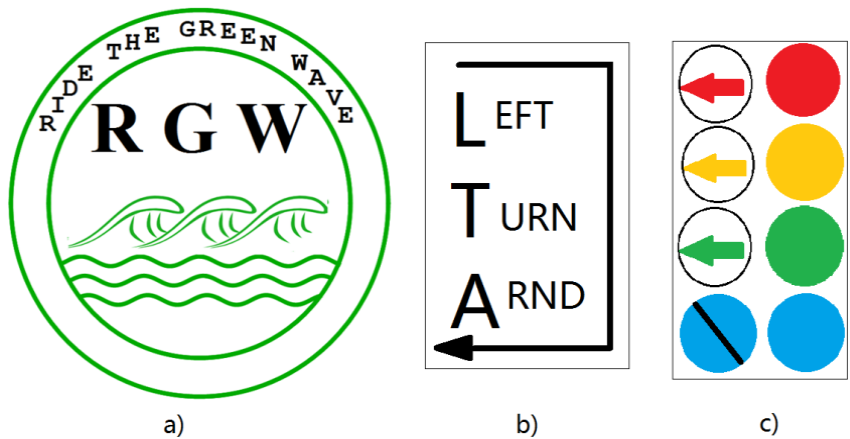

Figure 10. Signs and signals to support RGW-roads.

Figure 10 illustrates: a) sign designating RGW-road, b) sign indicating a left-turn-around, c) blue light indicates whether or not direct-left-turns are permitted.

## 3. Generalization of Cartesian grid to existing roads

Figure 11 shows a portion of a regular Cartesian road-network and a topologically-equivalent network.

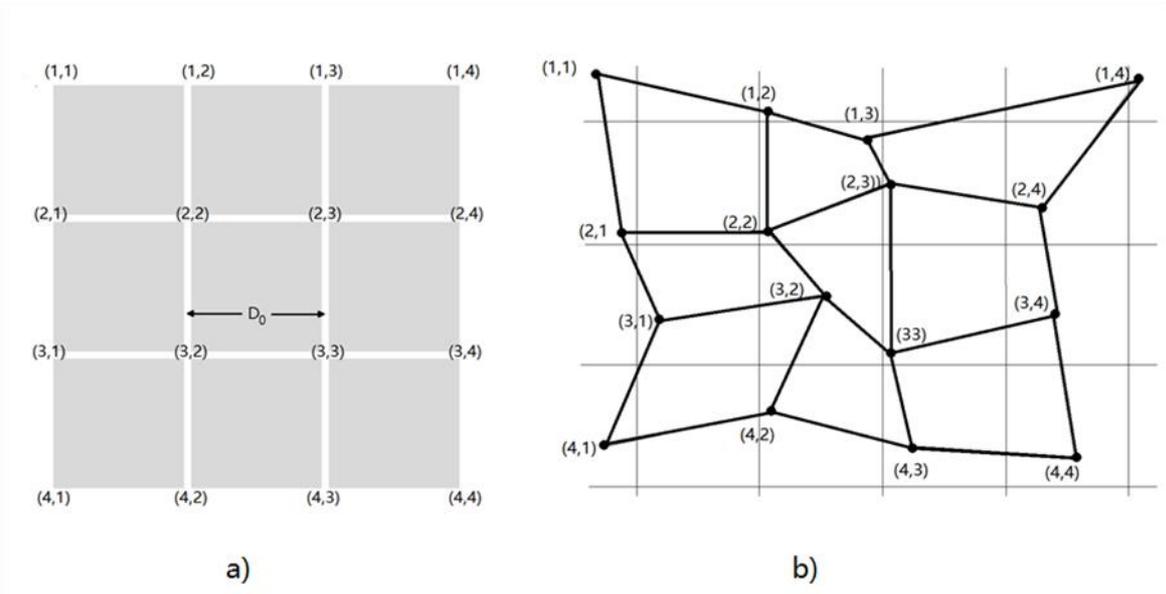

Figure 11. Generalizing Cartesian RGW results to topologically-equivalent road-networks. a) Cartesian network. b) Generalization of Cartesian network.

In Figure 11, a) illustrates a Cartesian road-network and b) illustrates a topologically-equivalent road-network.

Section 2 results were developed for a regular Cartesian road grid. This section generalizes Section 2 results to existing suburban arterial road-networks. An RGW-node in Figure 3 is the intersection of two RGW-roads. An *RGW-node*, on existing arterial roads, is defined as one of: 1) a signalized intersection used to control green-arrow motion, or 2) the locations of virtual traffic signals, not at intersections, used to control green-arrow motion. Virtual RGW-nodes which are not at intersections are introduced as a control device to slow down green-waves moving at exceedingly high speeds and to help coordinate traffic signals on RGW-road-networks

*Green-arrow laws appropriate for Figure 11b*. The first four green-arrow laws-of-motion given in Section 2 are applicable to Figure 11b) but the fifth law, which required green-waves move at a constant speed, needs modification. The modified fifth law: green-waves move with a speed equal to the distance between RGW-nodes divided by $T_g$. The modified fifth law implies green-waves move more quickly (slowly) than the green-wave speed in Figure 11a when the separation between RGW-nodes in Figure 11b is greater (less) than the separation between RGW-nodes in Figure 11a. Several results apply to Figure 11.

*Result 3.1*. Corresponding traffic signals in Figure 11a and 11b simultaneously turn green, amber and red.

*Result 3.2*. Green-arrow lengths (when the green-arrow is entirely within a block) in Figure 11b are equal to internodal-distance as in Figure 11a. The head and tail of a green-arrow move with the green-wave speed of the block they are in. Consequently, green-arrow length changes continuously as a green-arrow crosses an RGW-node with different green-wave speeds.

***Result 3.3***. Results 2.2 − 2.14 are valid for Figure 11b.

Figure 12, which shows portions of Figure 11b, illustrates how results from Figure 11b can be applied to existing roads.

***Result 3.4***. Results 3.1 – 3.3 are also valid when "Figure 11b" is replaced with "Figure 12".

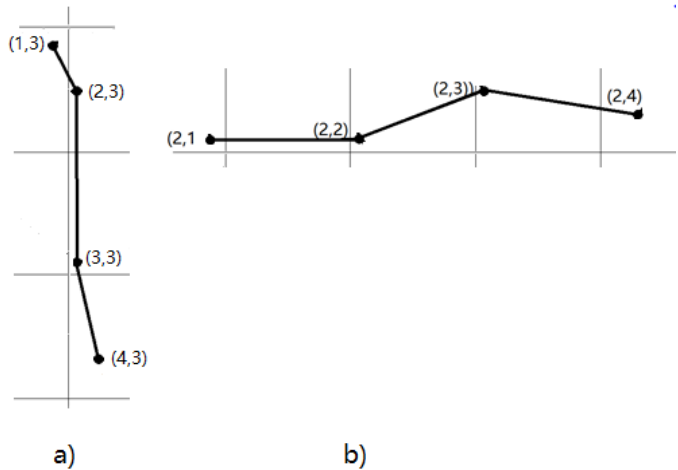

Figure 12. Representative north-south and east-west roads.

Define a *basic-road-network* as one where no more than two roads meet at an intersection. The discussion so far describes a mapping of green-arrow motion from a Cartesian road grid to an arbitrary basic-road-network, i.e., three RGW-roads cannot have a common intersection. In Figure 3, vehicles that travel at the green-wave speed make every traffic signal and this is easy for motorists to do since green-wave speed is constant in this figure. However, in Figures 11b and 12 green-wave speed depends on position which implies that for irregular road networks motorist speed guidance is needed.

For an arbitrary basic-road-network five problems are identified. 1) How will motorists know the green-wave speed at their location and how can they get in and stay in a green-wave? 2) When RGW-nodes are sufficiently far apart green-wave speeds may exceed the speed limit by an unacceptable amount. How can green-wave speed be kept within acceptable limits? 3) When RGW-nodes are sufficiently close together green-wave speed may be intolerably slow. How can green-wave speed be kept from being unacceptably slow? 4) Although left-turn-arounds (LTAs) allow motorists to make cloverleaf left-turns from RGW-roads, this is often less convenient than direct-left-turns. Can RGW-roads be designed so that when traffic flow is light motorists can make the more convenient direct left-turns? 5) Motorists may travel faster or slower than green waves. How will motorists be encouraged to travel at green-wave speeds? Answers to the identified problems are given below.

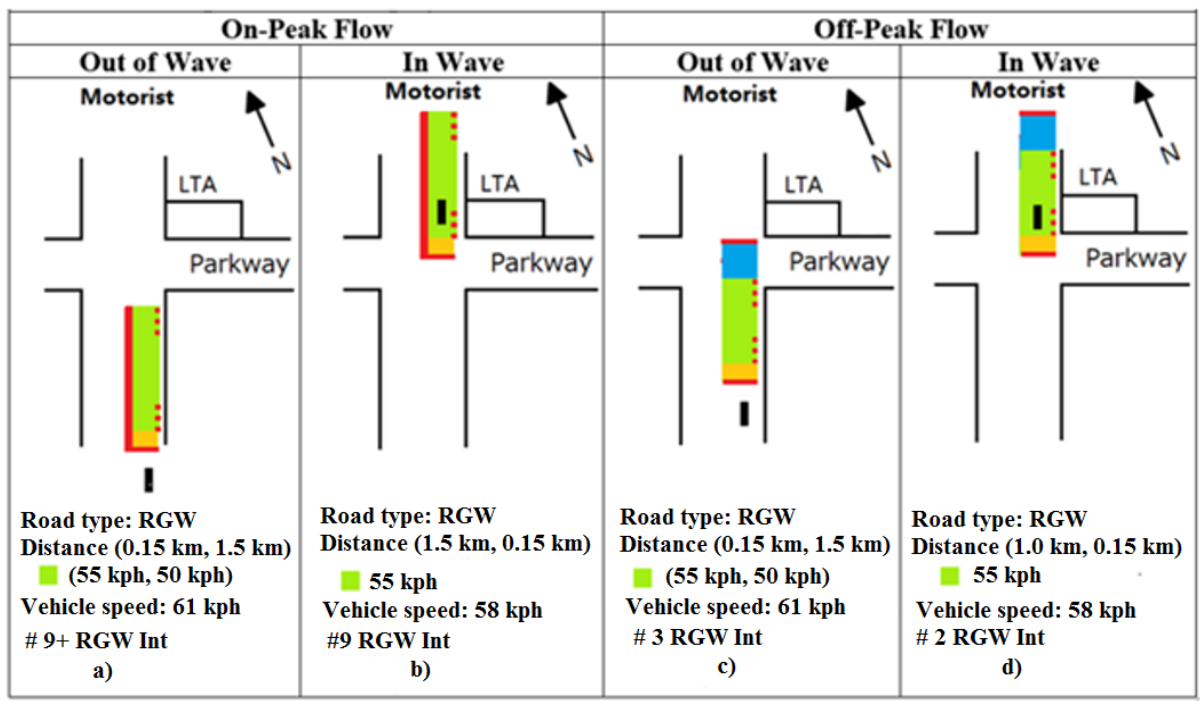

Figure 13. Representative road-to-traveler feedback device (RTFD) displays.

The RTFD, a smart phone application illustrated in Figure 13, solves the first problem by showing motorists their location and speed relative to a green-wave. The motorist who has the RTFD is represented by black rectangles. Green rectangles in a) and b) represent green-arrows. Green and blue rectangles in c) and d) represent left-turn-arrows (Figure 14). Solid vertical red lines in a) and b) indicate direct-left-turns are not permitted. Speeds in a) and c) preceded by a green square indicate green-wave speed, respectively, of the wave in front of and behind the motorist and in b) and d) indicate green-wave speed motorist is in. Distance in a) and c) indicate distance to the wave in front of and behind motorist. Distance in b) and d) represent distance to front and back of wave motorist is in. When traffic conditions permit, the RTFD guides motorists so they can get and stay in a green-wave.

The second problem is solved (see Section 4.2) by using virtual RGW-nodes (defined in the introduction). The third problem is most easily solved by judiciously selecting arterials to be converted to RGW-roads and the fourth problem is solved using the blue lights illustrated in Figure 10, the red lines in Figure 13 and the left-turn arrows illustrated in Figure 14.

The fifth problem deals with motorist driving behaviour. We consider different types of drivers: 1) speedy drivers who like to travel faster than the green-wave, 2) slow drivers who like to drive slower than the green-wave, 3) aggressive drivers who attempt to get in or stay in a green-wave and 4) self-driving vehicles. RGW-roads train fast (slow) drivers to slow down (speed up) with a mild penalty (wait at a red-light). Drivers who run red-lights to get in or stay in a green-wave are discouraged from doing that by receiving a moving-traffic violation. Although we anticipate RGW-road trains most drivers to travel at green-wave speeds, the deleterious effects of drivers that travel faster or slower than the green-wave speed is simulated in Section 5. Consistent with section 5 results, it is anticipated that connected and automated vehicles (CAV) can effectively use RGW-roads.

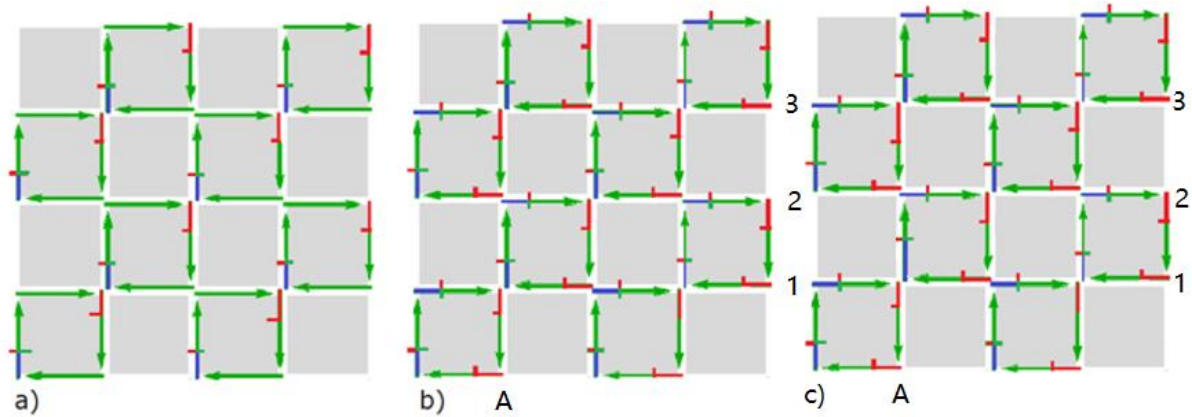

Figure 14. Maximum vehicle flow toward or away from a city during peak flow using left-turn-arrows and reduced flow arrows.

Left-turn- and reduced-flow-arrows, illustrated in Figure 14, respectively, have blue and red tails provide an alternative to cloverleaf left-turns during times of peak flow. These arrows have the same law of motion as green-arrows but have modified rules for changing traffic signals defined by how vehicles in these arrows move. Vehicles in the green portion of a left-turn arrow can go straight or make a right turn without having to stop; vehicles in the blue portion of a left-turn arrow have the additional capability of also being able to make a left-turn without stopping. Vehicles in the green portion of a reduced flow arrow who go straight make every traffic signal but those in the red portion get stopped at the first traffic signal they encounter.

Figure 14 a) illustrates maximum flow capacity in the east, west and north directions with direct left-turn capability for northbound traffic in the blue portion of a left-turn arrow. Maximum flow in the east and west directions is achieved using green-arrows. Maximum flow in the north direction is achieved using left-turn-arrows in the north direction and reduced-flow-arrows in the south direction. Here, flow in the southern direction has been sacrificed to allow direct left-turns for northbound vehicles. Part b) of the figure illustrates maximum flow capacity in the eastern and northern directions. Here vehicles in the blue portion of an arrow can make direct left-turns without a traffic stop. In part c) of the figure, part b) has been modified so that the northbound left-turn-arrow on road A between roads 1 and 2 has a lengthened blue portion and the southbound green-arrow on road A between roads 2 and 3 has a lengthened red portion. This satisfies an increased demand for left-turns on road A between roads 1 and 2.

*Left-turn-arrows*, illustrated in Figure 14 have blue tails, obey green-wave laws-of-motion but allow vehicles in the blue portion of the arrow to make left-turns without stopping for a traffic signal. *Reduced-capacity-arrows* illustrated in Figure 14 have red tails and are essential for left-turn-arrows to function properly. The figure shows how left-turn-arrows can be used to obtain maximum flow subject to constraints imposed by traffic flow demand. Red shaft portions of a green-arrow indicate flow in this direction has been sacrificed to allow direct left-turns in the opposite direction. In part a) of the figure this is possible because the red segment perpendicular to the arrow, at the beginning of the blue portion, turns traffic signals for southbound traffic red when it enters an intersection and similar comments could be made for parts b) and c) of this image. Left-turn-arrows allow maximum flow and direct-left-turns in selected directions at the expense of flow opposite the direction of the left-turn-arrow.

This paper describes a new class of roads called RGW-roads which are intermediate between freeways and customary arterial roads. On RGW-roads vehicles that travel at the recommended speed make all traffic signals and with left-turn-arounds simultaneously have maximum flow capacity. RGW-roads have the convenience of direct left-turns when traffic is light. Table 1 compares RGW- and existing-roads. The implementation of RGW roads, and RTFD require infrastructure investments.

**Table 1. Comparison of RGW-Roads with Streets, Arterial and Collector Roads***

| Road Class | Accessibility | Uninterrupted Flow | Cost/km | Speed |
|---|---|---|---|---|
| **Streets** | 4 | 1 | 1 | 1 |
| **Collector Roads** | 3 | 2 | 2 | 2 |
| **Arterial Roads** | 3 | 2 | 2 | 3 |
| **RGW-Roads** | 3 | 3 | 3 | 3 |
| **Freeways** | 1 | 4 | 4 | 4 |

**\* Numbers indicate attribute score from 1 to 4. Thus, streets are most accessible, freeways are the least accessible, arterial and RGW-roads are about equally accessible and are more accessible than freeways. RGW-0roads cost ore per km than arterial roads because of the need to construct cloverleaf left turns and implement RGW signals but cost less per km than freeways**

## 4. Case study

Section 2 developed the Uninterrupted Maximum Flow (UMAXFLOW) method for a regular arterial Cartesian road grid. In Section 3, the Section 2 method is extended so that it applies to a basic-road-network. Section 4 shows how to apply the method described in Section 3 to a representative suburban two-way arterial road. Section 4.1 describes the constrained minimization problem that could be used to select RGW-nodes and how RGW-nodes were actually selected for Telegraph Road (Alexandria, Virginia USA). Section 4.2 describe how to use the selected RGW-nodes to coordinate traffic signals.

**4.1 Placement of RGW-nodes**

Table 2 shows the location of traffic signals on Telegraph Road and the name of the cross roads/streets. This information was obtained from Google Maps. Figure 15 shows a graph that facilitates RGW-node selection. The abscissa shows the distance in kilometres along

Telegraph Road from Route 1. Blue dots in the figure correspond to existing traffic signals; amber dots corresponds to real and virtual RGW-nodes. Traffic signal numbers in Figure 15 correspond to physical traffic signals in Table 2. RGW-Node locations in Figure 15 correspond to Table 3 entries. Real and virtual RGW-nodes are easily distinguished in Figure 15: real RGW-nodes correspond to and are above traffic signals and this is not true for the virtual RGW-nodes. For example, RGW-nodes above traffic signals 1 and 4 are real and the node a little more than 1 km away from route 1 is virtual.

An optimisation procedure for RGW-node placement will now be described. Ideally, green-wave speed will be identical to the speed limit everywhere since this allows motorists to get to their destinations in the least amount of time and with the least attention to the RTFD. In general green-wave $v_g$ speed and the speed limit $v_{SpdLmt}$ (both piecewise continuous functions) will vary with position $o$ on an RGW-road. Define $\eta$ by

$$\eta^2 = \frac{1}{L} \int \left( v_g(o) - v_{SpdLmt}(o) \right)^2 do \tag{3}$$

Motivations for equation 3 are: 1) when green-wave speed $v_g$ is as constant as it can be, it will be easy for the driver to stay in a green-wave and, 2) if $v_{SpdLmt}$ is independent of position, then $\eta^2$ is the variance of $v_g$. Thus, $\eta$ defined by equation (3) is similar to the standard deviation of green-wave speed. The objective in RGW-node placement is to minimize (3) subject to the constraints

$$v_{min} \leq v_g(o) \leq v_{max} \tag{4}$$

$$\xi_i \leq \xi_{max,i}, \quad i = 1, 2, \ldots n \tag{5}$$

Constraint (4) is needed because vehicles should not go much faster than the speed limit (which determines $v_{max}$) and motorists lose interest in staying in a green-wave when $v_g$ is beneath some threshold (which determines $v_{min}$). Constraint (5) is needed for the $i^{th}$ location because Result 2.11 implies that when $\xi_i$ is above threshold value $\xi_{max,i}$ then $T_{gx}$ is too small to meet motorist cross flow demand. Note that constraint (5) restricts having cross-walks or intersections near midblock (see Results 2.6 and 3.3).

Appropriately choosing RGW-nodes is extremely important since it determines green-wave speed (see Results 3.4, 3.3, and 2.2) and the quality $\eta$ of the driving experience. In Section 4.2 it will become clear that once RGW-nodes and cycle time are chosen, traffic signal offsets are determined. The Mathematica RGW-TLP program automates the process. Result 2.3 shows that green-wave speed can be controlled using the cycle time, so by increasing the cycle time, green-wave speed can be decreased as illustrated in Figure 16. Decreasing green-wave speed is necessary at times when vehicle road density is high as required by Figure 1 and Equation (1).

**Table 2. Location of Signalized Intersections on Telegraph Road**

| # | Existing Traffic Signal Cross Streets | Odometer [km] |
|---|---|---|
| 1 | Route 1 | 0.000 |
| 2 | Belvoir Woods Parkway | 0.159 |
| 3 | Chynoweth Street | 1.962 |
| 4 | Lockport Place | 2.276 |
| 5 | Fairfax County S | 3.655 |
| 6 | Fairfax County N | 3.776 |
| 7 | Newington Road | 5.155 |
| 8 | Beulah Street | 6.236 |
| 9 | Hilltop Center Drive | 6.449 |
| 10 | Jeff Todd Way | 7.504 |
| 11 | Hayfield Road | 8.735 |
| 12 | Devereux Circle Drive | 10.667 |
| 13 | S. Van Dorn Street | 10.876 |
| 14 | S. Kings Highway | 11.191 |
| 15 | Rose Hill Drive | 12.332 |
| 16 | The Parkway | 14.231 |
| 17 | Franconia Road | 15.495 |
| 18 | Farmington Drive | 15.731 |
| 19 | Lenore Lane | 16.076 |
| 20 | N. Kings Highway | 16.180 |
| 21 | Huntington Avenue | 16.288 |

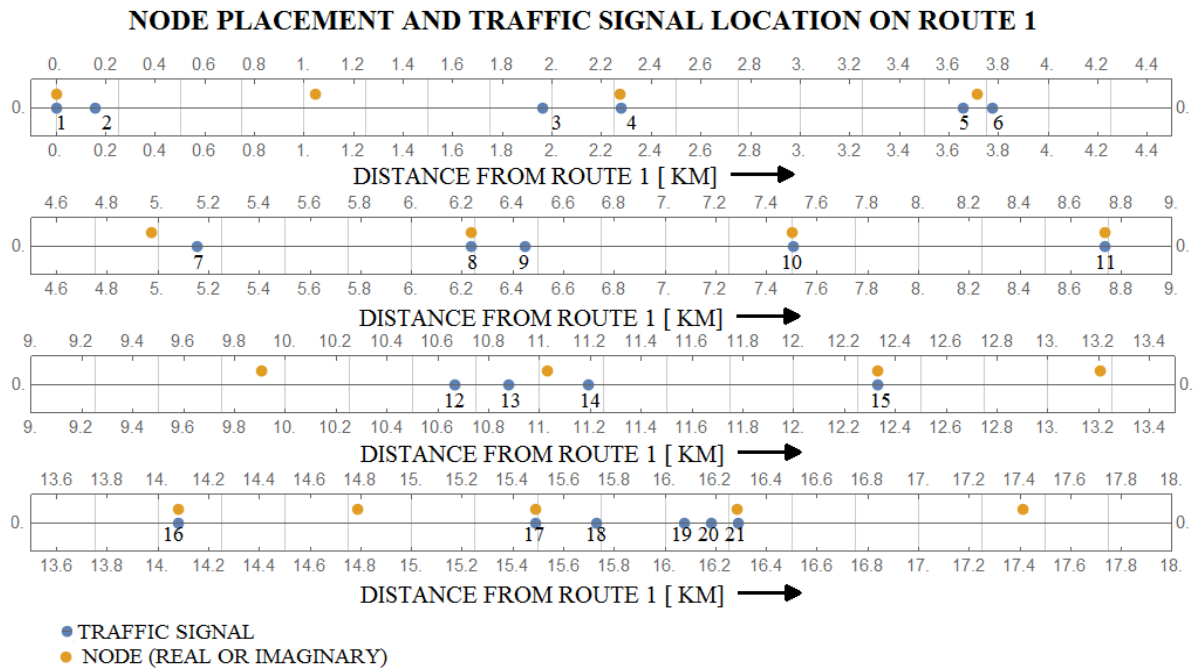

Figure 15. Traffic signal and node locations on Telegraph Road.

The procedure defined by equations (3), (4) and (5) was not used since $\xi_{max,i}$ parameters are not available and the program for doing the minimization procedure has not been written. Heuristic considerations for choosing RGW-node location in Figure 15 are now described. We want as many traffic signals as possible on Telegraph Road to be RGW-nodes since this allows maximum isotropic flow on Telegraph Road and the cross roads (Figures 3, 11, 12 and video 1). Typically, this is not possible because green-wave speed will be unacceptably slow (see Results 3.4, 3.3 and 2.2) when traffic signals are close together. We started by placing RGW-nodes at all traffic signals and then removed RGW-nodes at the less heavily travelled roads when green-wave speed is unacceptably slow. Traffic signal 1 in Figure 15 corresponds to Route 1, a heavily trafficked potential RGW-road and since it is highly desirable to make this intersection an RGW-node that was done. Similar comments apply to traffic signals 10, 11, 15, 16, 17 and 21. Traffic signals 5 and 6 are important and we need high flow at these locations. This is accomplished by placing a virtual RGW-node midway between these traffic signals. This allows reasonable time for cross flow without incurring an unacceptably slow green-wave speed. Similarly a virtual RGW-node was placed between signals 13 and 14. Traffic signals 20 and 21 correspond to heavily trafficked cross streets. Here instead of placing an RGW-node between these traffic signals, traffic signal 21 was made an RGW-node. This was done to increase green-wave speed between traffic signal 17 and 21. This placement allows maximum time for flow onto Telegraph Road at intersection 21 while allowing acceptable time for flow onto Telegraph Road at intersection 20. Virtual RGW-nodes are used to slow down green-wave speed when it is excessive. Examples are virtual RGW-nodes placed at locations near the 1, 9.9, 13.2 and 14.8 km markings. The RGW-node near the 5 km mark is an interesting case. We would have liked to make traffic signal 7 an RGW-node since doing that would allow maximum time for flow between Telegraph and Newington roads. However, making Newington road an RGW-node would have resulted in an excessive green-wave speed between the RGW-node between traffic signals 5 and 6 and Newington Road. Green-wave speed was reduced and an acceptable time for flow between Telegraph and Newington Road was obtained by placing an RGW-node near the 5 km mark.

**4.2 Traffic signal coordination on Telegraph Road**

Traffic signal timing in Table 3 was accomplished using the technique of Section 4.1 and principles that follow. A cycle time of two minutes is a short enough time so that vehicles do not overflow the Fairfax County exit ramps in Table 2 and is sufficiently long that the total time for clearing intersections is not excessive. Parameters $v_{max}$ and $v_{min}$ were, respectively, chosen to be nominally $v_{SpdLmt} \pm 16 \; km \; hr^{-1}$. Parameters $\xi_{max,i}$ are chosen to have the values given in Table 3. Principles for choosing node location are stated below.

1. Place an RGW-node at every intersection where a heavily trafficked or RGW-road intersects an RGW-road. As indicated in 2 and 3 below, this may need to be reversed when RGW-nodes are too close together or to satisfy constraints (4) and (5).

2. When heavily trafficked intersecting roads are close to one another either make the more heavily travelled road an RGW-node or place a virtual RGW-node between the two roads or place the RGW-node on the intersection that gives the most favourable green-wave speed. The RGW-node is placed so that both roads have an adequate value for $T_{gx}$.

3. Calculate green-wave speed $v_g$ and $\xi_i$ at different locations along the RGW-road using the program RGW-TLP which incorporates Results 2.2, 3.3 and Figure 4. Adjust RGW-nodes so that constraints (4) and (5) are satisfied.

4. The methodology described here works best on suburban roads where traffic signal density is not too high. If the objectives in node placement cannot be met with the assumed cycle time, investigate changing: 1) $v_{max}$ and/or $v_{min}$, 2) $\xi_{max,i}$ 3) cycle time. A working paper [23] shows that, with suitable road modification, principles summarized in Sections 2 and 3 can be satisfied with *urban traffic signal densities* on an RGW-road-network.

Table 3 applies the principles described above and the logic described in Section 4.1 to enable uninterrupted maximum flow on Telegraph Road.

The meaning of the different columns in Table 3 will now be described. Columns 1 and 2, respectively, identify row number and the name of the cross-street/road for a particular row. Entries in column 2 with parenthesis indicate the count of real and virtual RGW-nodes. All highlighted rows correspond to RGW-nodes. Those RGW-nodes which have names correspond to real nodes and are designated (N x) while those without names correspond to virtual nodes and are designated (V x). Rows that are not highlighted correspond to the intersection of RGW-roads with non-RGW-roads. Column 3 describes locations of items in column 2 with respect to Route 1. Column 4 indicates if a particular location is a node (real or virtual). Columns 5 and 6 describe the speed limit and length of green-wave, respectively. Columns 7, 8 and 9, respectively, describe green-wave speed, the dimensionless offset distance $\xi$ (Figure 4), and green-time in the forward direction (along Telegraph Road) $T_{gf}$. Columns 10, 11 and 12, respectively, describe green-time in the cross direction $T_{gx}$ (perpendicular to Telegraph Road), the offset-time $T_{offset}$ and the reduced-offset-time $T_{roffset}$. The reduced-offset-time is the onset time for a green signal on Telegraph Road.

A description of how to use Table 3 to time traffic signals follows. Calculations in Table 3 were made using a cycle time of 120 seconds. Column 12 tells when, in the 120 second cycle time, traffic signals first turn green. Thus, Belvoir Woods Pkwy turns green (allowing flow on Telegraph Road) at 110.9 sec. Column 9 indicates that the traffic signal at Belvoir Woods Pkwy for forward motion on Telegraph Road stays on for 78.3 sec. Column 10 indicates the traffic signal which allows flow across Telegraph Road is green for 41.7 sec. The green-times of 78.3 and 41.7 seconds include the amber and all red-time which are not specified. Observe that $110.9 + 78.3 + 41.7 = 230.9$ sec and Mod $[230.9, 120] = 110.9$ sec which implies the cycle repeats. Similar comments apply to every signalized intersection in Table 3. Observe that for every entry in Table 3, $T_{gf} + T_{gx} = 120\ s$

A description of how values in Table 3 are obtained follows. Columns 2 and 3 were obtained from Table 2. Values in column 4 were found using the procedures described in Sections 4.1. Speed limit information given in column 5 was obtained from a state of Virginia maintained web site (https://www.virginiaroads.org/maps/VDOT::vdot-speed-limits-map/about) Once node locations in column 4 are chosen, the remaining entries in Table 3 are determined and were found using equations in the RGW-TLP Mathematica program described below.

Values in columns 6 and 7 were obtained using Result 2.3. Values in columns 8, 9 and 10 were, respectively, obtained using Figure 4, Results 2.9 and 2.10. Here $T_{offset}$ in column 11 is the amount of time it takes the head of a green-wave to travel from Route 1 to the designated location. It is easily computed using green-wave speed $v_g$ and the distance between nodes $L_g$ obtained from column 6. Reduced-offset-times $T_{roffset}$ shown in column 12 are calculated differently for the case of nodes and for the case of non-RGW-roads intersecting RGW-roads. For nodes $T_{roffset} = \text{Mod}[T_{offset}, T_{CycleTime}]$. For non-RGW-roads intersecting RGW-roads values in column 12 are calculated using Results 2.11 or 2.12 depending on the $d$ parameter. The $d$ parameter is found using an algorithm that reproduces visual calculations exhibited in Figure 5. This paper does not discuss how long traffic signals are amber or all-red since these topics are discussed elsewhere [34 - 38]. Table 3 describes traffic signal parameters on Telegraph Road using a 2 min cycle time.

**Table** 3. **Coordination of traffic signals on Telegraph Road to get uninterrupted maximum stable flow.**

| 1 | 2 | 3 | 4 | 5 | 6 | 7 | 8 | 9 | 10 | 11 | 12 |
|---|---|---|---|---|---|---|---|---|---|---|---|
| = | Cross Street Nodes V & Re | Odometer [km] | RGW-Node? | Speed Limit [kph] | $L_g$ [km] | $v_g$ [kph] | $\xi$ [ ] | $T_{gf}$ [sec] | $T_{gx}$ [sec] | $T_{offset}$ [sec] | $T_{roffset}$ [sec] |
| 1 | Route 1 (N 1) | 0. | Yes | 72.4 | 1.046 | 62.8 | 0 | 60.0 | 60.0 | 0. | 0. |
| 2 | Belvoir Woods Pkwy | 0.159 | No | 72.4 | 1.046 | 62.8 | 0.1523 | 78.3 | 41.7 | | 110.9 |
| 3 | Node (V 2) | 1.046 | Yes | 72.4 | 1.229 | 73.7 | 0 | 60.0 | 60.0 | 60. | 60. |
| 4 | Chynoweth St | 1.961 | No | 72.4 | 1.229 | 73.7 | 0.2552 | 90.6 | 29.4 | | 104.7 |
| 5 | Lockport Place (N 3) | 2.275 | Yes | 72.4 | 1.438 | 86.2 | 0 | 60.0 | 60.0 | 120. | 0. |
| 6 | Fairfax County S | 3.652 | No | 72.4 | 1.438 | 86.2 | 0.0425 | 65.1 | 54.9 | | 57.4 |
| 7 | Node (V 4) | 3.714 | Yes | 72.4 | 1.258 | 75.5 | 0 | 60.0 | 60.0 | 180. | 60. |
| 8 | Fairfax County N | 3.775 | No | 72.4 | 1.258 | 75.5 | 0.0486 | 65.8 | 54.2 | | 57.1 |
| 9 | Node (V 5) | 4.972 | Yes | 72.4 | 1.263 | 75.8 | 0 | 60.0 | 60.0 | 240. | 0. |
| 10 | Newington Rd | 5.154 | No | 72.4 | 1.263 | 75.8 | 0.1439 | 77.3 | 42.7 | | 111.4 |
| 11 | Beulah St (N 6) | 6.235 | Yes | 64.4 | 1.268 | 76.1 | 0 | 60.0 | 60.0 | 300. | 60. |
| 12 | Hilltop Center Dr | 6.447 | No | 64.4 | 1.268 | 76.1 | 0.1675 | 80.1 | 39.9 | | 50. |
| 13 | Jeff Todd Way (N 7) | 7.503 | Yes | 64.4 | 1.231 | 73.9 | 0 | 60.0 | 60.0 | 360. | 0. |
| 14 | Hayfield Rd (N 8) | 8.734 | Yes | 56.3 | 1.171 | 70.3 | 0 | 60.0 | 60.0 | 420. | 60. |
| 15 | Node (V 9) | 9.905 | Yes | 56.3 | 1.126 | 67.6 | 0 | 60.0 | 60.0 | 480. | 0. |
| 16 | Devereux Cir Dr | 10.664 | No | 56.3 | 1.126 | 67.6 | 0.3257 | 99.1 | 20.9 | | 40.5 |
| 17 | S Van Dorn St | 10.874 | No | 56.3 | 1.126 | 67.6 | 0.14 | 76.8 | 43.2 | | 51.6 |
| 18 | Node (V 10) | 11.031 | Yes | 56.3 | 1.298 | 77.9 | 0 | 60.0 | 60.0 | 540. | 60. |
| 19 | S Kings Hwy | 11.189 | No | 56.3 | 1.298 | 77.9 | 0.1214 | 74.6 | 45.4 | | 52.7 |
| 20 | Rose Hill Dr (N 11) | 12.33 | Yes | 56.3 | 0.874 | 52.5 | 0 | 60.0 | 60.0 | 600. | 0. |
| 21 | Node (V 12) | 13.204 | Yes | 56.3 | 0.874 | 52.5 | 0 | 60.0 | 60.0 | 660. | 60. |
| 22 | The Parkway (N 13) | 14.079 | Yes | 56.3 | 0.706 | 42.3 | 0 | 60.0 | 60.0 | 720. | 0. |
| 23 | Node (V 14) | 14.785 | Yes | 56.3 | 0.706 | 42.3 | 0 | 60.0 | 60.0 | 780. | 60. |
| 24 | Franconia Rd (N 15) | 15.491 | Yes | 56.3 | 0.793 | 47.6 | 0 | 60.0 | 60.0 | 840. | 0. |
| 25 | Farmington Dr | 15.728 | No | 56.3 | 0.793 | 47.6 | 0.2982 | 95.8 | 24.2 | | 102.1 |
| 26 | Lenore Ln | 16.072 | No | 56.3 | 0.793 | 47.6 | 0.2677 | 92.1 | 27.9 | | 43.9 |
| 27 | N Kings Hwy | 16.177 | No | 56.3 | 0.793 | 47.6 | 0.1359 | 76.3 | 43.7 | | 51.8 |
| 28 | Huntington Ave (N 16) | 16.285 | Yes | 56.3 | 1.126 | 67.6 | 0 | 60.0 | 60.0 | 900. | 60. |
| 29 | Node (V 17) | 17.411 | Yes | 56.3 | 1.126 | 67.6 | 0 | 60.0 | 60.0 | 960. | 0. |

Table 3 was constructed using our judgment for $\xi_{max,i}$ values.

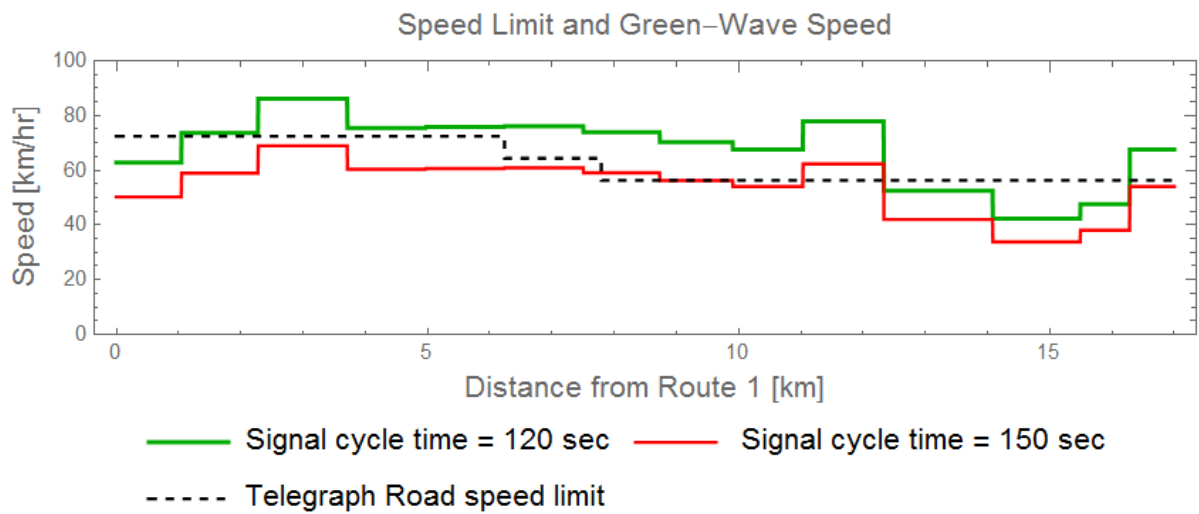

Figure 16. Speed limit and green-wave speed on Telegraph Road for different cycle times.

Figure 16 shows the speed limit on Telegraph Road and green-wave speeds with 120 and 150 s cycle time. As indicated by Figure 1, as vehicle road density increases, green-wave speed needs to decrease so vehicles can keep up with the green-wave. Figure 16 shows this can easily be done by increasing the cycle time. An alternate method for reducing green-wave speed is to adjust node locations. Figure 16 shows green-arrow heads have abrupt speed changes as they cross certain nodes which implies infinite acceleration for the green-arrow head. This is not a problem for vehicles that want to maintain their position in the green-wave since they can accelerate gradually to regain their position in the green-wave before it gets to the next traffic signal.

*Result 4.1.* When an RGW road is exclusively occupied by dingoes (defined in Table 4) driven vehicles separated by a constant headway with one northbound source and one northbound sink, the traffic light coordination described in Table 3 implies a vehicle flow rate independent of position. The same result is true if *north* in the above statement is replaced with *south*, or *east* or *west*.

*Why result is true.* 1) Green-wave speed $v_g$ is proportional to the distance between nodes. 2) A constant headway implies vehicle density $\rho$ is inversely proportional to the distance between nodes. 3) Flow rate $q = \rho\, v_g$. Observations 1) – 3) imply truth of result. The validity of Result 4.1 is confirmed by Table 3: observe that $v_g/L_g = 60.0$ independent of the chosen row.

## 5. RGW-SIM Description, Applications and Use

RGW-SIM (Ride the Green Wave Simulation) is Mathematica code that was developed to 1) validate equations developed in this paper and 2) to estimate how well the proposed method for developing progressions on Telegraph Road will work when it is populated with automated vehicles, ideal and heterogeneous drivers. Sections 5.1, 5.2 and Appendix A describe how RGW-SIM was used to achieve the first and second goals.

**5.1 Validation of equations used to develop Table 3.**

This section and Appendix A describe how RGW-SIM was used to show that Table 3 and the equations used to derive Table 3 are correct.

*Description of RGW-SIM*.

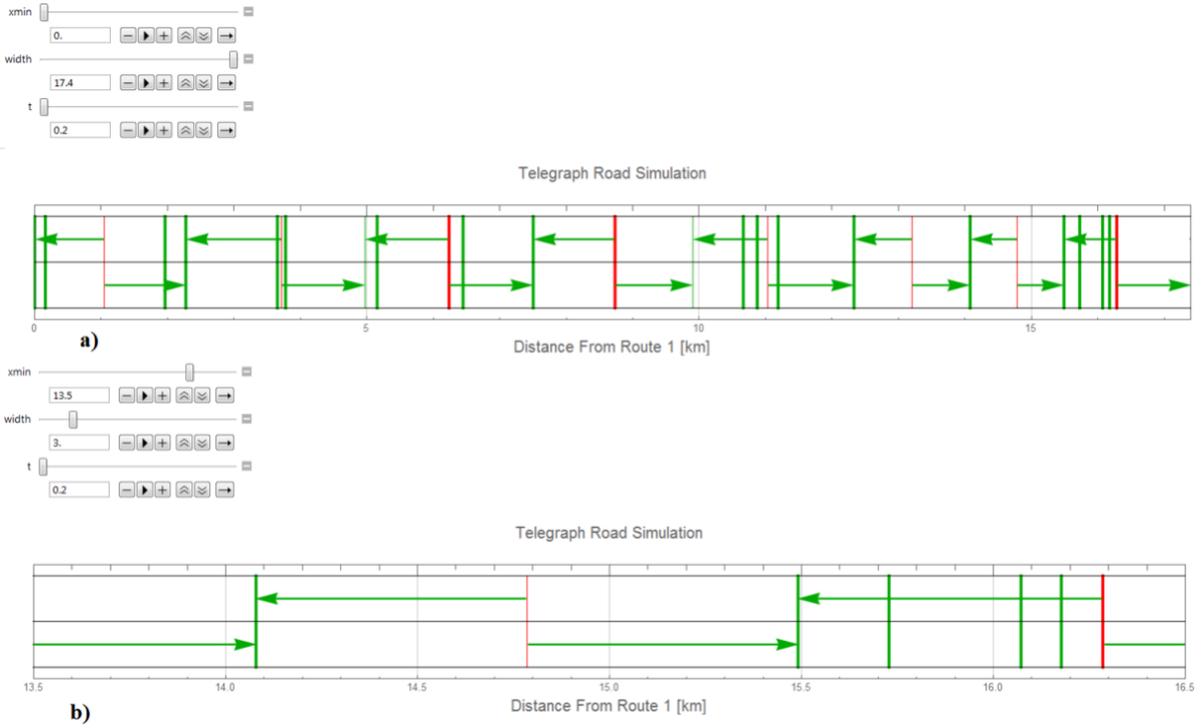

Figure 17. RGW-SIM output display without vehicles. a) Display showing the entirety of Telegraph Road at $t = 0.2\ s$ and b) display at $t = 0.2\ s$ showing details at the northern portion of Telegraph Road.

Representative RGW-SIM output displays are shown in Figure 17. The slider labelled "width" controls how much of Telegraph Road is shown; the slider labelled "xmin" determines what portion of the road is shown and the slider labelled "t" shows the road state at time t. The heavy red and green vertical lines represent existing traffic signals and the colour (green, yellow, red) indicates the state of those traffic signals. Light red and green vertical lines correspond to virtual nodes and their colour represents the state of those virtual traffic signals. By moving the time slider to the right the user can rapidly advance time and by repeatedly clicking on the plus sign underneath the time slider the user can advance time as slowly as desired.

RGW-SIM needs the head and tail positions of north and southbound arrows as a function of time. The column labelled $v_g$ in Table 3, gives green-wave speed as a piecewise function of position. Integrating $v_g$ yields (6) and (7) which give the position of the north and southbound green-arrow head positions as functions of time. Here $xGWN[t]$ and $xGWS[t]$ are head positions, respectively, of a northbound and a southbound green-arrow.

$$\mathrm{xGWN}[t] = \begin{cases} 0, & t \leq 0, \\ d_i + v_{g,i}[t - (i-1)T_g], & (i-1)T_g \leq t \leq iT_g,\ i=1,\ldots N-1, \\ d_N, & t \geq NT_g, \end{cases} \qquad (6)$$

$$\text{xGWS}[t] = \begin{cases} d_N, & t \leq 0, \\ d_{N-i+1} - v_{g,N-i}[t-(i-1)T_g], & (i-1)T_g \leq t \leq iT_g, \ i=1,\ldots,N-1, \\ 0, & t \geq NT_g. \end{cases} \quad (7)$$

Here, $N$ is the number of RGW-nodes and is 17 for the case study, $d_i$ is the odometer reading of RGW-node $i$, $L_{g,i}$ is the length of green-wave traveling from RGW-node $i$ to $i+1$, and $v_{g,i}$ is the speed of the green-wave traveling from RGW-node $i$ to $i+1$. The parameter $T_g$ is the time traffic signal is green, amber and all-red on the RGW-road and equals 60 s for the case study (Section 4). The parameter $d_i$ is shown in column 3 of Table 3. For the case study $d_0 = 0, d_1 = 1.046, \ldots, d_{17} = 17.411$. The green-wave speed in the northbound direction from RGW-node $i$ to RGW-node $i+1$ is given by

$$v_{g,i} = \frac{d_{i+1} - d_i}{T_g} = \frac{L_{g,i}}{T_g}, \quad i = 1, \ldots, N-1$$

For the case study, green-wave lengths $L_{g,i}$ and green-wave speeds $v_{g,i}$ are given in column 6 and 7, respectively, of Table 3. Expanded versions of (6) and (7) appropriate to the case study are given in Figure A1.

In (6) and (7) $xGWN[t]$ and $xGWS[t]$ give the head position of northbound and southbound green-arrows as a function of time. Since red and green traffic signals are on for 60 seconds the tail of the zeroth northbound green-arrow as a function of time is given by $xGWN[t-60]$. Other northbound green-arrow head positions in the simulation are given by $xGWN[t-120\,i]$ where $i$ is a positive or negative integer. A similar procedure was used to generate southbound arrows.

*Conclusion.* The following discussion shows that RGW-SIM supports the validity of Table 3 and the equations used to derive Table 3. By advancing time, green-arrows move forward as specified by column 7 in Table 3 and as required by Result 2.3 and its generalizations. As illustrated in Figure A3, RGW-SIM determines the states of real and virtual RGW-nodes using results given in columns 9 and 12 in Table 3 which are obtained using Results 2.9 - 2.14 and its generalizations. As time advances, green-arrows in Figure 17 advance, and RGW-SIM shows that traffic signals at real and virtual RGW-nodes turn green as arrows enter the intersection and turn amber and then red as they leave the intersection. The observation that green-arrow motion is in agreement with traffic signal behaviour, even though green-waves and traffic signal state are described by different equations (see Figure A1 for green-arrow movement and Figures A2 and A3 for traffic signal behaviour), confirms that Table 3 and the equations used to create that table are correct.

**5.2 Dependence of traffic performance parameters on driver characteristics.**

The claim has been made that vehicles traveling north or south on Telegraph Road at the RTFD-recommended speed make all traffic lights. RGW-SIM simulations support this claim and describe how the driving experience degrades when the road is populated by drivers who do not follow RTFD recommendations. This section describes the meaning of Table 5, how the values in the table were determined, and the reasonableness of the results.

**Table 4. Driver attributes used in RGW-SIM**

| Driver Type | Characteristics |
|---|---|
| All drivers | Telegraph Road is modelled as a two-lane bi-directional road. No passing is allowed. All vehicles obey traffic light signals. When released from a traffic signal all drivers (except dingoes) accelerate at a constant rate of 2.5 m per s$^2$ until they reach their desired speed. When speed limit changes, with the exception of dingoes, all vehicles adapt to the new speed with accelerations/decelerations of 2.5 m per s$^2$. |
| Dingo | Dingoes travel at green-wave speed $v_g$ when road is exclusively populated with dingoes. Whenever green-wave speed changes dingo acceleration and deceleration is determined by $x2[t]$ (Figure 19). |
| Wolf | Wolves travel at speed limit when road is exclusively populated with wolves. |
| Cheetah | Cheetahs travel 15 km/hr above the speed limit when road is exclusively populated with cheetahs. |
| Tortoise | Tortoises travel 15 km/hr below the speed limit when road is exclusively populated with tortoises. |
| Dingo-Wolf | Wolves travel at the speed limit except when they are constrained to go slower by a dingo driven vehicle ahead of them. Dingoes travel at green-wave speed when they are the lead vehicle and go at speed of vehicle ahead of them otherwise. |
| Dingo-Cheetah | Cheetahs travel 15 km/hr above the speed limit except when they are constrained to go slower by a dingo driven vehicle ahead of them. Dingoes travel at green-wave speed when they are the lead vehicle and go at speed of vehicle ahead of them otherwise. |
| Dingo-Tortoise | Tortoises travel 15 km/hr below the speed limit. Dingoes travel at green-wave speed when they are the lead vehicle and go at speed of vehicle ahead of them otherwise. |

Dingo-like drivers can either be thought of as ideal human drivers or as connected and automated vehicles that strictly follow rules specified in Table 4. Other types of drivers (cheetahs, tortoises and wolves) represent driving behaviours that substantially deviate from RTFD recommendations.

*Modelled scenario*. Several simplifying assumptions were made in RGW-SIM to estimate how well the proposed method for coordinating traffic signals will work. 1) Although much of Telegraph Road has one lane in each direction, a small portion of the road has three lanes in each direction and a substantial portion has two lanes in each direction. For simplification, Telegraph Road is modelled as having one lane in each direction without passing. 2) Although vehicles typically get on and off Telegraph Road at many traffic signal controlled intersections, RGW-SIM works by assuming all northbound vehicles originate at the Route 1 intersection and exit at the Huntington Ave intersection, and that all southbound vehicles originate at the Huntington Ave intersection and exit at the Route 1 intersection. 3) Although drivers have a continuum of driving characteristics whose behaviours are more accurately described as random variables drawn from the continuum, RGW-SIM simplifies drivers as described by Table 4. 4) Although Telegraph Road has occasional gentle changes in direction and mild inclines, RGW-SIM ignores these features. RGW-SIM does not simulate Telegraph Road per se; rather, it simulates an imaginary two-lane bidirectional road with traffic signals located at the same locations as those on Telegraph Road and with the same speed limits.

*Explanation of Table 5 columns*. Columns 2 - 5 describe the mixture of dingoes, wolves, cheetahs and tortoises used in the simulation. The headway of one second in simulation 3 implies that this simulation represents what might be achieved by a computer-driven vehicle using an RTFD. A headway of two seconds used in all other simulations can be achieved by human-driven vehicles. Column 6 gives the mean travel time [s] to get from intersection of Route 1 and Telegraph Road to crossing the Telegraph-Huntington intersection. Column 7 gives the mean number of red light stops per passenger car unit (pcu), a parameter of interest for pollution control. Column 8 describes the mean time spent waiting at traffic lights, a parameter of interest to motorists and traffic engineers. The difference between the mean travel time and 907 seconds is reported in column 9 and the value in column 9 divided by 907 seconds expressed as a percentage is reported in column 10. The maximum pcu flow rate per lane is reported in column 11. This is defined as the number of vehicles that cross the Huntington Avenue intersection (the finish line) divided by time $\Delta t$ where $\Delta t$ is defined as the difference in time between the last and first vehicle to cross the Huntington-Telegraph Road intersection. Because isotropic flow is modelled in the simulation, the mean pcu rate is (where given), neglecting amber and red clearance time, half the maximum pcu rate. Some entries in column 12 are left blank because the relationship between mean and maximum pcu rate is not known. Further details on how values in Table 5 are calculated are given in Section 3 of Appendix A.

*Simulation description*. All simulations were calculated for northbound vehicles and were done for vehicles in the first northbound green-wave (the one not yet visible but about to enter Telegraph Road at location 0 in Figure 17). Southbound green-waves were not populated with vehicles because it is expected that northbound and southbound simulations will yield almost identical results and calculations done for dingoes show this to be true. Only one northbound green-wave was populated with vehicles (we could have populated all the northbound green-waves shown in Figure 17 with vehicles) because creating time-space functions such as that shown in Figure A.6 is time consuming and doing so can exceed available computer resources. Because only one green-wave is populated, red font values in Table 5 are steady-state results deduced from transient simulations.

*Table 5 font colour*. Simulation 1 was done with only the first northbound green-wave having 27 vehicles, each separated with a 2 second headway to get the results shown in Table 5. Simulation 1 was done again with all northbound green-waves populated with 27 vehicles. Dingoes are described by a single time-space function $x2[t]$ (exhibited in Figure 19) which makes it feasible to populate all northbound green-waves with vehicles. It was found that for simulations 1, 2 and 3 steady-state and transient results shown in Table 5 are identical. This result, observed in simulations 1, 2 and 3, is predicted by Result 5.1.

**Result 5.1.** If vehicles entering an RGW-road are in a single –wave and they stay in that green-wave until they exit the RGW-road then transient and steady-state performance are identical.
*Why result is true*. The condition that a certain driving behaviour results in vehicles stay in one green-wave implies that, given the same driving behaviour, vehicles will stay in every green-wave and vehicle performance in each green-wave will be identical. Thus, traffic performance calculated for one platoon of vehicles is identical to that for all platoons of vehicles.

In the fifth simulation, 27 cheetah-like drivers entered Telegraph Road in the first green-wave and stayed in that green-wave till they crossed the finish line (Telegraph-Huntington intersection). Result 5.1 asserts that the steady-state and transient results are identical for this case and normal and bold font are used to represent the results from simulation 5. All normal font values in Table 5 have been explained.

Results 5.2 and 5.3 imply red font values in Table 5 are steady-state lower bounds.

**Result 5.2.** If vehicles entering an RGW-road are in a single –wave exit the RGW-road in multiple waves then transient results are calculated for those vehicles.
*Why result is true*. Vehicles which populate the next green-wave will also spread out into more than one green-wave as time progresses and vehicles from the two waves will both be in a single green-wave.

**Result 5.3.** Column 6 – 9 results in Table 5 cannot decrease with an increase in traffic.
*Why result is true*. 1) Drivers try to minimize the parameters calculated in columns 6 – 9 but are sometimes prevented from achieving this by other drivers on the road. 2) Result 5.2 implies that in the steady-state more vehicles will be in each wave. Observations 1 and 2 imply the result.

Blue font results in Table 5 are transient results because transient simulations were done. Bold values in column 11 were calculated from the transient simulation but are steady-state values because of Result 5.1. Bold values in column 12 have half the values in column 11 because neglecting the all red and yellow portion of the cycle the appropriate denominator for column 12 is half that for column 11.

**Table 5. Dependence of Telegraph Rd Steady-State Performance on Driver Attributes**

| 1 | 2 | 3 | 4 | 5 | 6 | 7 | 8 | 9 | 10 | 11 | 12 |
|---|---|---|---|---|---|---|---|---|---|---|---|
| Simulation # | Dingo | Wolf | Cheetah | Tortoise | Mean Travel Time Averaged Over PCUs | Mean Red Light Stops per pcu | Mean Time Waiting at Red Light Stop | Mean Delay Time per pcu | % Change Mean Delay Time per pcu | Maximum pcu flow rate | Mean pcu flow rate |
| | % | % | % | % | $[s]$ | $[\,]$ | $[s]$ | $[s]$ | $[\,]$ | $\left[\frac{v}{hr\,ln}\right]$ | $\left[\frac{v}{hr\,ln}\right]$ |
| | | | | | MTT | MRLS | MRLST | MDT | | | |
| 1 | 100 | 0 | 0 | 0 | 907 | 0 | 0 | 0 | --- | **1800** | **900** |
| 2 | 100 | 0 | 0 | 0 | 907 | 0 | 0 | 0 | --- | **1080** | **540** |
| 3 | 100 | 0 | 0 | 0 | 907 | 0 | 0 | 0 | --- | **3600** | **1800** |
| 4 | 0 | 100 | 0 | 0 | *1089* | *2.8* | *95* | *182* | *20.0* | 608 | --- |
| 5 | 0 | 0 | 100 | 0 | 897 | 5 | 80. | −10 | -1.1 | **1800** | **900** |
| 6 | 0 | 0 | 0 | 100 | *1554* | *5.6* | *231* | *647* | *71.3* | 403 | --- |
| 7 | 77.8 | 22.2 | 0 | 0 | *1081* | *3.3* | *91* | *174* | *19.2* | 338 | --- |
| 8 | 77.8 | 0 | 22.2 | 0 | *916* | *0.22* | *9* | *9* | *0.99* | 797 | --- |
| 9 | 77.8 | 0 | 0 | 22.2 | *1389* | *4.4* | *177* | *482* | *53* | 132 | --- |

- In simulations 1 to 3 all green-waves are initially filled to capacity and no vehicles are outside of a green-wave. In simulations 4 to 9 all vehicles are initially in first green-wave and rest of road is devoid of vehicles.
- Two seconds headway and 27 vehicles for every simulation except simulation 3, which has a one second headway and 54 vehicle in each green wave.
- In simulation 2 the number of vehicles in each green-wave is populated according to a Poisson distribution.
- Maximum pcu flow rate is defined as number of vehicles (27 or 54) that cross Huntington Ave intersection divided by the time it takes those vehicles to cross intersection. This is the flow rate during the time traffic signal is green. For simulations 1-3, and 5 the average flow rate is half the maximum flow rate.
- Numbers in *red italic* are lower bounds. **Bold numbers** indicate the transient and steady-state values are identical. Blue font numbers represent transient results. It is not known if blue font numbers are steady-state upper or lower bounds. If one accepts the model used, normal and bold font numbers are correct steady-state calculations.

*Analysis of Table 5 results.* Next, we examine Table 5 results. The mean travel time to get from Route 1 to just past Huntington Avenue is smaller in simulations 1-3 than for every simulation except simulation 5. On the average, simulation 5 does the trip 10 seconds faster than simulations 1-3 because cheetahs traverse the last leg of the journey faster than dingoes.

Notice that cheetahs paid a price for this 10 second average decrease in journey time: the cheetahs stopped for a mean of 5 traffic signals and on average spent 80 seconds waiting at traffic signals. Observe that in simulations 1-3, dingoes make every traffic light. This is expected, since traffic signal timing and dingo motion was designed to obtain this result. In simulation 4 wolves are stopped for an average of 2.8 traffic signals but tortoises are stopped for an average of 5.6 traffic signals. For wolves, the green-wave speed is sometimes higher and sometimes lower than the speed limit but tortoises typically travel slower than the speed limit, so it is reasonable that the mean number of traffic signals where tortoises are stopped is greater than that of wolves. Note that in simulation 6 the mean time spent waiting for a traffic signal to change to green is 231 seconds and 95 seconds is spent waiting for a traffic signal to change to green in simulation 4. It is reasonable that wolves spend less time than tortoises waiting for a red traffic signal to turn green. The reason: tortoises go so slow they repeatedly are left behind by the green-wave and then have to wait almost the entire red light time for the traffic signal they missed. On the other hand, when wolves go quicker than the green-wave they are typically only a little ahead of the green-wave behind them and only have to wait a short time for that green-wave to release them from the red light. We conclude that the results from simulations 1-9 are reasonable.

The Table 5 results do not take into account the statistical distribution of arrival and reaction times, acceleration/deceleration capabilities etc. Although our simulation results address simplified scenarios, they have significant implications for real-life implementation of the proposed methodology. Field experimentation coupled with simulation could address these issues. However, field experimentation is outside the scope of this paper. Simulations described in Table 5 investigate how variations of individual vehicle speed from green-wave speed influence the total time to traverse Telegraph Road (column 6), the mean number of traffic lights stopped at (column 7), the mean time spent waiting at red traffic lights (column 8), the mean delay time per pcu (column 9) and the maximum and average flow rates (columns 11 and 12). Sources and destinations in our simulation are the two ends of Telegraph Road. This is a gross simplification but having multiple sources and destinations is outside the scope of this paper. Although there are locations on Telegraph Road with one, two and three lanes in each direction, for simplicity Telegraph Road is simulated as having a single lane in each direction.

## 5.3 Description of RGW-SIM Model

This section describes the model used to produce Table 5. A mathematical description of how dingoes move with green waves without having the delta function acceleration of green-waves as they cross nodes where green-wave speed changes abruptly is given in Section 5.3.1. Appendix A describes several features of RGW-SIM. 1) The motion of green-waves and the turning on and off of green, amber and red traffic signals. 2) To satisfy the requirements of Table 4 vehicles need to speed up, and slow down. Vehicles also need to slow down to stop for traffic signals. 3) Details for how vehicles were controlled so that they satisfied Table 4 requirements.

*5.3.1 Modelling dingo motion across speed-change nodes.* Observe from Table 3 that the head and tail speed of a green-wave change abruptly as green waves traverse a node (real or virtual) whenever the green-wave speeds on either side of the node are different. Nodes where green-wave speed is different on different sides of a node are termed *speed-change nodes*. Thus the head and tail of green-waves have infinite acceleration the instant they cross a speed-change node. Of course, dingo driven vehicles trying to travel at green-wave speed cannot experience infinite acceleration.

Here we describe how we model dingo driven vehicles so they move with the green-wave but without infinite acceleration. Impose the condition that the spatial $x$ coordinate of dingo driven vehicles satisfy the differential equation

$$\frac{d^2 x}{d t^2} + \gamma \frac{dx}{dt} + \omega_0^2 x = g(t), \quad \gamma = 2\,\omega_0 \tag{8}$$

where

$$g(t) = \begin{cases} \omega_0^2\, xGWN[t] & \text{for northbound dingo driven vehicles} \\ \omega_0^2\, xGWS[t] & \text{for southbound dingo driven vehicles} \end{cases} \tag{9}$$

and $\gamma$ or $\omega_0$ are parameters chosen to have the desired acceleration.

The condition $\gamma = 2\,\omega_0$ makes (8) the equation of a critically damped oscillator [40]. The critically damped oscillator has one free parameter which we take as $\gamma$. As shown in Figure 18, by choosing $\gamma = 0.5$ and $\omega_0 = 0.25$, dingo driven vehicles have an acceleration of less than 1.7 m s$^{-2}$ when crossing speed-changing nodes, which is well within the capabilities of modern vehicles. Had we required vehicles to move at green-wave speed vehicle acceleration would be described by delta functions instead of the realizable accelerations shown in Figure 18. Other traffic flow applications of differential equations include the intelligent driver model [41] and fluid flow models [42, 43].

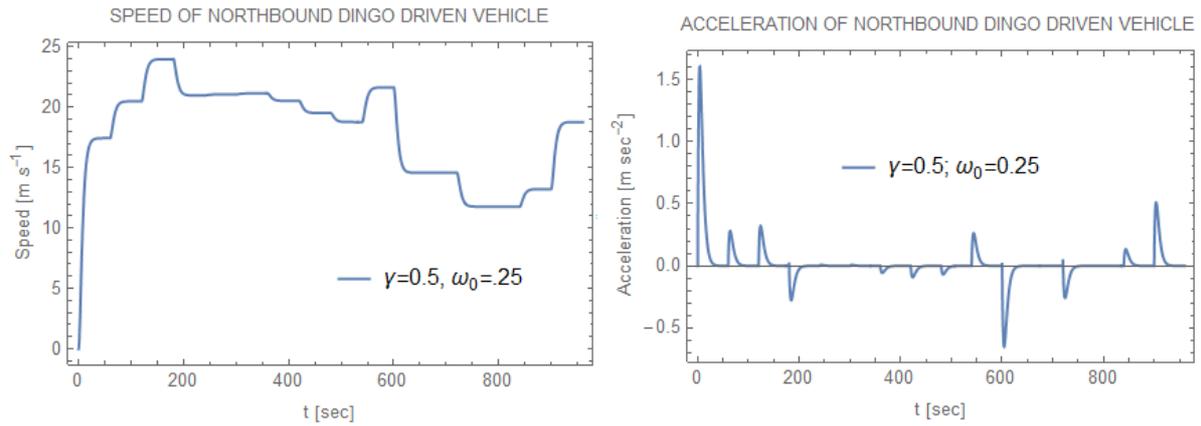

Figure 18. Dingo speed and acceleration calculated from (8) and (9).

The method used to generate Figure 18 is described in Figure 19.

```
In[52]:= sol = NDSolve[{x''[t] + .5 x'[t] + 0.0625 x[t] == 0.0625 xGWN[t],
         x[0] == 0, x'[0] == 0}, x, {t, -900, 3×960}];

In[53]:= x2[t_] = x[t] /. sol[[1]]; (* Solution with ω₀=0.25 and γ=0.5 *)

In[54]:= g1 = Plot[1000 x2'[t], {t, 0, 960}, Frame → True,
         FrameLabel → {{"Speed [ m s⁻¹]", ""},
           {"t [sec]", "SPEED OF NORTHBOUND DINGO DRIVEN VEHICLE"}}];

In[55]:= g2 = Plot[1000 x2''[t], {t, 0, 960}, Frame → True, PlotRange → {-1, 1.7},
         FrameLabel → {{"Speed [ m s⁻¹]", ""},
           {"t [sec]", "ACCELERATION OF NORTHBOUND DINGO DRIVEN VEHICLE"}}];
```

Figure 19. Mathematica code used to generate Figure 18.

The argument of NDSolve in line 52 is equation (8). The factor $\omega_0^2$ multiplying $xGWN[t]$ is needed because when $x''[t]$ and $x'[t]$ are zero we want $x[t]$ to equal $xGWN[t]$ and also $x[t]$ must be dimensionally equal to $xGWN[t]$. The solution in line 52 is an interpolating function. Line 53 allows the interpolating function to be more easily manipulated and lines 54 and 55 essentially produce the graphs shown in Figure 18. Additional information on RGW-SIM is given in Appendix A.

## 6. Discussion

The requirement that this paper be of reasonable length necessitated limiting the paper's scope. Since details of how the RTFD works have been described in a patent application [17, 18] they were omitted from the paper. Proofs of many results stated in Sections 2 and 3 are given in reference [23] and are the subject of a future paper. Implicit in our discussion is the assumption that drivers will do their best to get in their desired position in a green-wave because that will get them to their destination in the least time with the fewest traffic stops. This appears not be an issue for CAVs though it has yet to be demonstrated.

The program (RGW-TLP) takes as input information given in columns 2, 3 and 4 in Table 3 and outputs results shown in column 5 – 12.  Choosing RGW-nodes cleverly is the hard part of getting optimum results out of RGW-TLP.  In the future we hope to develop a program that, taking input from a traffic engineer, automates RGW-node placement.

We would have liked to model the performance metrics shown in Table 5 with performance metrics obtained using current traffic signal timing.  This was not possible because actual traffic signal timing on Telegraph Road is unknown to us.  Also, the presence of induction coils indicates traffic signal timing on Telegraph Road is not deterministic whereas traffic signal timing in this work is deterministic.

In Table 3 some green-wave speeds (col 7) are greater than the speed limit (col 5), which suggests the approach taken in this paper is impractical.  This is not the case as indicated by the following argument. a) Green-wave speed can be decreased so it is beneath the speed limit at all points along Telegraph Road by increasing cycle time as indicated in Figure 16.
b) By appropriately placing virtual nodes, green-wave speed can be adjusted to be equal to or less than the speed limit.  Table 3 was constructed with the view that green-wave speed brackets the speed limit. c) When traffic permits, drivers on Telegraph Road typically go between 8 and 16 km/hr (5 and 10 mph) over the speed limit.  Thus, green-wave speeds in Table 3 are consistent with typical driving behaviour.   This is not inconsistent with traffic engineering practice.

Characteristics of two-dimensional vehicle movement [39] were not discussed in this paper. However, we do address an important concern of reference [39]: "Intersections are the critical nodes of the urban traffic network, and their design has great implications for traffic safety and efficiency".  As illustrated in Figures 3, 5-7, 15 and video-1, this paper achieves maximum efficiency at real RGW-nodes by enabling maximum flow through intersections.

## 7. Conclusions

The most important contribution of this paper is the concept of RGW-roads which is compared in Table 1 with existing streets/roads/freeways.  On an RGW-road, vehicles traveling at the recommended speed make every traffic signal on that road regardless of how far they travel. It is anticipated that although the speed limit on RGW-roads will be slower than highway or freeway speed limits, the cost to build RGW-roads will be substantially less than that of highways or freeways because there is no need for over/under passes.  This paper describes, using the Section 4 case study, how almost any arterial road can be converted to an RGW-road by providing suitable traffic signal timing.

Typically, vehicles on RGW-roads adjust their speed to achieve non-stop/uninterruptable flow and are advised, at all times, of the speed they should travel by the RTFD, a smart-phone device (Figure 13). The use of the RTFD to guide motorist speed is the second most important contribution of this paper. It is expected to have greater effectiveness in achieving uninterrupted flow in conjunction with CAVs. In earlier unpublished papers and presentations we have discussed how results described here for a single road can be extended to a network of intersecting roads [23, 33].

The third most important contribution of this paper is a different perspective on how to achieve progressions. The usual approach is to use programs that time traffic signals so that vehicles going at a constant speed will have maximum bandwidth and make traffic signals over a finite range. In the usual approach, bandwidth is lost as progression length increases. In the approach used here maximum bandwidth is achieved over the entire length of the road and progressions can be arbitrarily long because our approach makes use of the RTFD and virtual RGW-nodes. Our approach focuses on the optimum method for moving green-arrows, which correspond to vehicle platoons, and traffic signal timing follows directly from the movement of green-arrows.

This paper also develops a new notation, illustrated in Figure 14, which shows how the movement of green or left-turn arrows control traffic signals. The use of one-way streets when combined with left-turn arrows allows traffic engineers to use green-waves, originally designed for isotropic flow, to accommodate existing non-isotropic flow demands.

How to get uninterruptable flow on a Cartesian road network is known. Section 3 describes how to extend these results to arbitrary road network, topologically equivalent to a Cartesian road network. This is new and is made possible by the concept of virtual nodes, the RTFD, and Results 2.x and Results 3.x.

RGW-roads have the maximum possible potential vehicle flow, because aside from traffic signal all-red time, green-waves are always moving through every RGW-intersection (see Figure 1) and, with automated vehicles, can do so at speeds approximating the speed limit.

At periods of maximum flow in four orthogonal directions, here called north, south, east and west, direct left-turns are not possible without sacrificing maximum flow. The limitation of not being able to make direct left-turns during periods of maximum flow is overcome with cloverleaf left-turns (Figure 9). When maximum flow is not needed, motorists can make the more convenient direct left-turns. Since motorists can make direct left-turns when maximum flow is not needed, but cannot make direct left-turns when maximum flow is needed, blue traffic signals (Figure 10) tell motorists without an RTFD when they can and cannot make direct left-turns. The adoption of left-turn-arounds, commonly used on freeways, to RGW-roads to facilitate potential uninterrupted maximum flow on RGW-roads is a new application of this well-known concept.

When traffic flow in one direction is maximum but traffic flow in the opposite direction is substantially less than maximum, left-turn arrows (Figure 14) allow motorists properly positioned in the left-turn arrow to make direct left-turns on two-way roads without stopping.

RGW-roads require road to vehicle communications, which is accomplished by the RTFD (Figure 13) phone application, but do not require vehicle-to-vehicle or vehicle-to-road communications. Thus, implementing RGW-roads is substantially easier than implementing road systems that require vehicle-to-vehicle and vehicle-to-road communications, e.g. CAV systems. Most of the simulations reported in Table 5 had two seconds headway between vehicles which can easily be handled with an RTFD alone. Simulation three had one second headway between vehicles which we believe can safely be done with an appropriate CAV system. Thus, RGW-roads can work with just the RTFD phone application alone but are expected to work even better when CAV systems are implemented.

The program RGW-TLP used to produce the Telegraph Road progressions and the program RGW-SIM were developed for this paper and can be freely downloaded. The development of RGW-TLP was facilitated by Results 2.x and Results 3.x for $x = 1, 2, \ldots$ . These results, the RGW-road concept, our use of left-turn-arounds to facilitate maximum flow on RGW-roads, the visual arrow language introduced in Figure 14, the use of blue traffic signals for left-turns, and the sign informing motorists they are on an RGW-road illustrated in Figure 14 are all original.

The typical commute to and from work is stressful. Driving is much less tension producing on RGW-roads where drivers are informed of the speed they should travel and where traffic lights are programmed to turn green as drivers approach them,. Thus, RGW-roads have the capability of substantially improving the quality of commuters' commutes.

The uninterrupted maximum stable flow approach described here can improve driver satisfaction and rush-hour traffic flow by giving motorists an ability to drive to and from work with minimal/no traffic signal stops. Corresponding improvements are expected for public transport operations. RGW-roads and RGW-road-networks can offer lower capital-cost alternatives to highways/freeways/expressways, can have an impact on road-network design, and can reduce energy consumption and pollutant emissions by reducing vehicle traffic stops.

## Conflicts of Interest

The authors declare that there are no conflicts of interest regarding the publication of this article.

## Funding Statement

This project is self-funded.

## Acknowledgments

We thank Elise Miller-Hooks from the George Mason University Department of Civil, Environmental and Infrastructure Engineering department for suggesting Nathan Gartner to join our team. We thank David Friedman for finding the Virginia speed limit link given in Section 4.2 and Sara Friedman for preliminary experiments in finding the distance between traffic signals given in Table 3.

## Supplementary Materials

Video-1 and one page summary of invention https://ott.gmu.edu/wp-content/uploads/2020/06/GMU-20-016_Fact-Sheet_Green-Waves-on-Red-Sea.pdf can be downloaded. The programs RGW-TLP and RGW-SIM , referred to in this paper, can be downloaded.  The RGW-TLP file includes data to produce Table 3 and the RGW-SIM file include input data for the simulations reported here.

**APPENDIX A – DESCRIPTION OF RGW-SIM**

The overarching objective of RGW-SIM is to estimate how heterogeneous drivers and automated vehicles influence the performance of RGW-roads. Section 1 of this appendix describes how Table 3 results were validated. A description of how vehicles move is given in section 2 and section 3 describes the methodology used to calculate Table 5 results.

*1. Implementing green-wave motion and traffic signal behaviour.* This section describes how green-arrows move and how traffic signals are coordinated. The coordinated movement of north and southbound green-waves with the traffic signal behaviour validates Table 3 and the equations used to create that table. The position for the head of a single northbound green-arrow as a function of time is given by $xGWN[t]$ and since the cycle time is 120 seconds the tail of that arrow is given by $xGWN[t-60]$. This is illustrated in Figure A1.

$$xGWN[t\_] := \begin{cases} 0 & t < 0 \\ 0 + \frac{1.046}{60} t & 0 \le t < 60 \\ 1.046 + \frac{1.229}{60}(t-60) & 60 \le t < 120 \\ 2.275 + \frac{1.438}{60}(t-120) & 120 \le t < 180 \\ 3.714 + \frac{1.258}{60}(t-180) & 180 \le t < 240 \\ 4.972 + \frac{1.263}{60}(t-240) & 240 \le t < 300 \\ 6.235 + \frac{1.268}{60}(t-300) & 300 \le t < 360 \\ 7.503 + \frac{1.231}{60}(t-360) & 360 \le t < 420 \\ 8.734 + \frac{1.171}{60}(t-420) & 420 \le t < 480 \\ 9.905 + \frac{1.126}{60}(t-480) & 480 \le t < 540 \\ 11.031 + \frac{1.298}{60}(t-540) & 540 \le t < 600 \\ 12.33 + \frac{0.874}{60}(t-600) & 600 \le t < 660 \\ 13.204 + \frac{0.874}{60}(t-660) & 660 \le t < 720 \\ 14.079 + \frac{0.706}{60}(t-720) & 720 \le t < 780 \\ 14.785 + \frac{0.706}{60}(t-780) & 780 \le t < 840 \\ 15.491 + \frac{0.793}{60}(t-840) & 840 \le t \le 900 \\ 16.285 + \frac{1.126}{60}(t-900) & 900 \le t \le 960 \\ 17.411 & 960 \le t \end{cases}$$

a)

$$xGWS[t\_] := \begin{cases} 17.411 & t \le 0 \\ 17.411 - \frac{1.126}{60}(t-0) & 0 \le t < 60 \\ 16.285 - \frac{0.793}{60}(t-60) & 60 \le t < 120 \\ 15.491 - \frac{0.706}{60}(t-120) & 120 \le t < 180 \\ 14.785 - \frac{0.706}{60}(t-180) & 180 \le t < 240 \\ 14.079 - \frac{0.874}{60}(t-240) & 240 \le t < 300 \\ 13.204 - \frac{0.874}{60}(t-300) & 300 \le t < 360 \\ 12.333 - \frac{1.298}{60}(t-360) & 360 \le t < 420 \\ 11.031 - \frac{1.126}{60}(t-420) & 420 \le t < 480 \\ 9.905 - \frac{1.171}{60}(t-480) & 480 \le t < 540 \\ 8.734 - \frac{1.231}{60}(t-540) & 540 \le t < 600 \\ 7.503 - \frac{1.268}{60}(t-600) & 600 \le t < 660 \\ 6.235 - \frac{1.263}{60}(t-660) & 660 \le t < 720 \\ 4.972 - \frac{1.258}{60}(t-720) & 720 \le t < 780 \\ 3.714 - \frac{1.438}{60}(t-780) & 780 \le t < 840 \\ 2.275 - \frac{1.229}{60}(t-840) & 840 \le t < 900 \\ 1.046 - \frac{1.046}{60}(t-900) & 900 \le t < 960 \\ 0 & 960 \le t \end{cases}$$

b)

```
In[6]:= NBArrow[xTail_, xHead_] := Arrow[{{xTail, 0.3}, {xHead, 0.3}}]   (* Northbound Arrow *)

       NBArrow[xGWN[t - 300], xGWN[t - 240]],
       NBArrow[xGWN[t - 180], xGWN[t - 120]],
    →  NBArrow[xGWN[t - 60], xGWN[t]],
       NBArrow[xGWN[t + 60], xGWN[t + 120]],
       NBArrow[xGWN[t + 180], xGWN[t + 240]],
```

c)

Figure A1. Method for producing the moving green-arrows shown in Figure 17. a) Explicit representation of (6). b) Explicit representation of (7). c) The method for moving northbound green-arrows.

Line 6 defines the head and tail position of a northbound green-arrow. The command with the red arrow defines a green-arrow which at time $t = 0$ has a head that is about to enter Telegraph Road where it intersects route 1. Green arrow heads are separated from each other by 120 seconds and in each case green-arrows are 60 seconds long. Northbound green-arrow head motion is controlled by $xGWN[t]$ defined by (6). Similarly, $xGWS[t]$ defined by (7) describes the movement of southbound arrow heads.

Figure A2 shows the subroutine used to control traffic signal state in the simulation.

```
TLCntrl[t_, Troffset_, Tgf_] :=
 Module[{GreenStart, AmberStart, RedStart, temp1, temp2, temp3, temp4, output},
  GreenStart = Table[Troffset + 120 i, {i, -15, 15}];
  (* This went from -12 to 12 *)

  (* AmberStart = GreenStart + Tgf -6 ;
  RedStart = GreenStart + Tgf-1; *)

  temp1 = Select[GreenStart, # ≤ t &] // Last;  (* temp1 is green start time *)
  temp2 = temp1 + Tgf - 6; (* amber start time associated with green
   start time *)
  temp3 = temp1 + Tgf - 1; (* red start time associated with green start time *)
  temp4 = temp1 + 120 ; (* is next green start time *)

  output = { DGreen                              temp1 ≤ t < temp2
             Blend[{Yellow, Orange}, 1/3]        temp2 ≤ t < temp3 ;
             Red                                 temp3 ≤ t < temp4
             "This should not happen"            True           }

  output ]
```

Figure A2. Subroutine TLCntrl is used to control traffic signals in RGW-SIM.

TLCntrl arguments $t, T_{roffset}$ and $T_{gf}$, respectively, correspond to time [s] and the values given in columns 12 and 9 in Table 3. Examples of calls to TLCntrl are exhibited in Figure A3.

```
Thicker, TLCntrl[t, 0, 60], MLine[0] ,
Thicker, TLCntrl[t, 110.9, 78.3], MLine[0.159],
Thinner, TLCntrl[t, 60, 60], MLine[1.046],
Thicker, TLCntrl[t, 104.7, 90.6], MLine[1.961],
Thicker, TLCntrl[t, 0, 60], MLine[2.275],
Thicker, TLCntrl[t, 57.4, 65.1], MLine[3.652],
Thinner, TLCntrl[t, 60, 60], MLine[3.714],
Thicker, TLCntrl[t, 57.1, 65.8], MLine[3.775],
Thinner, TLCntrl[t, 0, 60], MLine[4.972],
```

Figure A3. Example of traffic signal calls in RGW-SIM

In the lines of Figure A3, observe that the second and third arguments of TLCntrl correspond to Route 1, Belvoir Woods Pkwy, …, Node (V 5) in Table 3 and that virtual and real RGW-nodes have the prefix "Thinner" and "Thicker", respectively, corresponding to the thin and thick traffic signal indicators shown in Figure 17.

*2 Implementing vehicle movement into simulation.* The laws of motion dictated by Table 4 imply that wolves, cheetahs and tortoises traveling on Telegraph Road either exclusively or with dingoes have to occasionally stop for traffic signals and since the speed limit on Telegraph Road varies with position, vehicles need to also slow-down and speed-up. Functions which describe how vehicles speed-up and slow-down are exhibited in Figure

a)
```
xSpeedUp[t_, X0_, V0_, V1_, A_, t0_] :=
    (* t is time [s] *)
    (* X0 is vehicle position at time t0 [m] *)
    (* V0 is vehicle speed [m s⁻¹] for time ≤ t0 *)
    (* V1 is vehicle target speed [m s⁻¹] *)
    (* A is vehicle positive acceleration [m s⁻²] *)
    (* t0 is time [s] when speed up begins *)
```

$$\begin{cases} V0\,(t-t0) + X0 & t < t0 \\ \frac{1}{2} A\,(t-t0)^2 + (t-t0)\,V0 + X0 & t0 \le t \le \frac{A\,t0 - V0 + V1}{A} \\ -\frac{V0^2 - 2\,V0\,V1 + V1^2 - 2\,A\,(t\,V1 - t0\,V1 + X0)}{2\,A} & t > \frac{A\,t0 - V0 + V1}{A} \end{cases}$$

b)
```
xSlowDown[t_, X0_, V0_, V1_, A_, t0_] :=
    (* t is time [s] *)
    (* X0 is vehicle position at time t0 [m] *)
    (* V0 is vehicle speed [m s⁻¹] for time ≤ t0 *)
    (* V1 is vehicle target speed [m s⁻¹] *)
    (* A is vehicle negative acceleration [m s⁻²] *)
    (* t0 is time [s] when speed slow down begins *)
```

$$\begin{cases} V0\,(t-t0) + X0 & t < t0 \\ \frac{1}{2} A\,(t-t0)^2 + (t-t0)\,V0 + X0 & t0 \le t \le \frac{A\,t0 - V0 + V1}{A} \\ -\frac{V0^2 - 2\,V0\,V1 + V1^2 - 2\,A\,(t\,V1 - t0\,V1 + X0)}{2\,A} & t > \frac{A\,t0 - V0 + V1}{A} \end{cases}$$

Figure A4. *x*SpeedUp and *x*SlowDown functions.

Observe that the codes and the arguments for xSpeedUp and xSlowDown are identical and that the meaning of the arguments are described as comments within the function. When

xSpeedUp (xSlowDown) is called the acceleration A is positive (negative) and $V1$ is larger (smaller) than $V0$. Figure A4 a) describes the coordinate of a vehicle speeding up from position $X0$ at time $t0$ with initial speed $V0$ to a final speed $V1$ with constant acceleration $A$. Similar comments apply to Figure A4 b).

Figure A5 describes a subroutine that outputs the position of a vehicle which at time $t0$ is at location $X0$ traveling with speed $V0$ and comes to a complete stop in a distance $d$ with constant deceleration.

```
SlowDownToStop[t_, t0_, x0_, v0_, d_] := Module[{a, T, temp, dstop},
  (* t is time [s] *)
  (* t0 is initial time [s] *)
  (* x0 is initial position [m] *)
  (* v0 is initial speed [m s^-1] *)
  (* d is stopping distance [m] *)
  a = - 1/2 v0^2/d;  (* needed deceleration [m s^-2] *)
  T = - v0/a;  (* time to stop [s] *)
  dstop = - 1/2 v0^2/a;  (* stopping distance [m] *)
  temp = { 1/2 a (t - t0)^2 + v0 (t - t0) + x0    t0 ≤ t < t0 + T
           x0 - 1/2 v0^2/a                         t0 + T ≤ t  };
  (* temp is position [m] as function of time [s] *)
  temp ]
```

Figure A5. Subroutine **SlowDownToStop** describes the position of a vehicle stopping for a traffic signal.

Figure A6 shows subroutines exhibited in Figures A4 and A5 which allow the RGW-SIM user to suitably control the motion of vehicles so that they obey all traffic signals and also obey rules of motion described in Table 4. Nominally ten such functions are required for each green-wave. Properly defining each time-space function in Figure A6 is a time-consuming task.

```
DT1P1[t_] := x2[t]

DT1P2[t_] :=
  ⎧ 0.001 xSpeedUp[t, 0, 0, 15.94, 2.5, 0]                        0 ≤ t < 237 - 10
  ⎪ 0.001 SlowDownToStop[t, 227, 3567.6, 15.94, 84.4]             227 ≤ t < 300 - 10
  ⎪ 0.001 xSpeedUp[t, 3652, 0, 15.94, 2.5, 290]                   290 ≤ t < 465.4 - 10
  ⎪ 0.001 xSlowDown[t, 6237.6, 15.94, 13.72, -2.5, 455.4]         455.4 ≤ t < 545 - 10
  ⎨ 0.001 SlowDownToStop[t, 535, 7330.7, 13.72, 172]              535 ≤ t < 602 - 10
  ⎪ 0.001 xSpeedUp[t, 7503, 0, 13.72, 2.5, 592]                   592 ≤ t < 625.8 - 10
  ⎪ 0.001 xSlowDown[t, 7791.8, 13.72, 11.47, -2.5, 615.8]         615.8 ≤ t < 864 - 10
  ⎪ 0.001 SlowDownToStop[t, 854, 10525, 11.47, 139]               854 ≤ t < 884 - 10
  ⎪ 0.001 xSpeedUp[t, 10664, 0, 11.47, 2.5, 874]                  874 ≤ t < 1020 - 10
  ⎪ 0.001 SlowDownToStop[t, 1010, 12197.6, 11.47, 132.4]          1010 ≤ t < 1082 - 10
  ⎩ 0.001 xSpeedUp[t, 12330, 0, 11.47, 2.5, 1072]                 1072 ≤ t

DT1P3[t_] := DT1P2[t]

                ⎧ DT1P2[t]                                             0 ≤ t < 485 - 20
                ⎪ 0.001 SlowDownToStop[t, 465, 6370.3, 13.72, 76.7]    465 ≤ t < 532 - 20
DT1P4[t_] :=    ⎨ x2[t - 194]                                          512 ≤ t < 577 - 20
                ⎪ 0.001 SlowDownToStop[t, 552, 7397.2, 13.72, 105.8]   557 ≤ t < 616 - 20
                ⎩ DT1P3[t]                                             596 ≤ t

                ⎧ DT1P4[t]                                             0 ≤ t < 714 - 22
                ⎪ 0.001 SlowDownToStop[t, 692, 8666.8, 11.47, 67.2]    692 ≤ t < 783 - 22
                ⎪ x2[t - 333]                                          761 ≤ t < 875 - 22
DT1P5[t_] :=    ⎨ 0.001 SlowDownToStop[t, 853, 10505.5, 11.47, 158.5]  853 ≤ t < 899 - 22
                ⎪ DT1P4[t]                                             877 ≤ t < 1434 - 22
                ⎪ 0.001 SlowDownToStop[t, 1412, 16203.5, 11.47, 81.5]  1412 ≤ t < 1502 - 22
                ⎩ 0.001 xSpeedUp[t, 16285, 0, 18.78, 2.5, 1480]        1480 ≤ t
```

Figure A6. Dingo-tortoise time-space functions. Use of functions **xSpeedUp** and subroutine **SlowDownToStop** to control the motion of a mixture of dingo-like and tortoise-like drivers traveling so as to obey traffic signals and rules of motion specified in Table 4.

Figure A6 describes how routines described in Figures A4 and A5 are utilized to control the motion of individual or groups of drivers. The method for controlling the motion of individual drivers in a graphics program is shown in Figure A7

```
Red, PointSize[.004],
Point[{DT1P1[t], .3}]],    (*1*)
Point[{DT1P1[t - 2], .3}]],  (* 2*)
Point[{DT1P1[t - 4], .3}]],  (* 3*)
Point[{DT1P1[t - 6], .3}]],  (* 4*)
Point[{DT1P1[t - 8], .3}]],  (* 5*)

Black, PointSize[.006],
 Point[{DT1P2[t - 10], .3}]],  (* 6 *)

Red, PointSize[.004],
Point[{DT1P3[t - 12], .3}]],  (* 7 *)
Point[{DT1P3[t - 14], .3}]],  (* 8 *)
Point[{DT1P3[t - 16], .3}]] ,  (* 9 *)

Black, PointSize[.006],
Point[{DT1P3[t - 18], .3}]],  (* 10 *)
```

Figure A7.  Graphics functions which control the motion of dingoes and tortoises.

Figure A7 shows that the first five vehicles are controlled by the function DT1P1 and Figure A6 shows that DT1P1 is the function $x2[t]$ which is a dingo interpolation function and is the solution of the differential equation shown in Figure 19.  Figure A8 shows that to obtain the mixture of vehicles driven by dingo-like and tortoise-like drivers, a random draw was used.  The same draw was used for simulations 7, 8 and 9 so that Table 5 results are not influenced by the randomness of the draw.

```
In[28]:= SeedRandom[987]; temp1 = RandomVariate[BernoulliDistribution[.2], 27]
Out[28]= {0, 0, 0, 0, 0, 1, 0, 0, 0, 1, 0, 0, 0, 0, 0, 0, 0, 1, 1, 0, 0, 1, 0, 0, 1, 0, 0}

In[29]:= temp2 = Range[27]
Out[29]= {1, 2, 3, 4, 5, 6, 7, 8, 9, 10, 11, 12, 13,
       14, 15, 16, 17, 18, 19, 20, 21, 22, 23, 24, 25, 26, 27}

In[30]:= temp3 = {temp2, temp1}ᵀ
Out[30]= {{1, 0}, {2, 0}, {3, 0}, {4, 0}, {5, 0}, {6, 1}, {7, 0}, {8, 0}, {9, 0}, {10, 1},
       {11, 0}, {12, 0}, {13, 0}, {14, 0}, {15, 0}, {16, 0}, {17, 0}, {18, 1}, {19, 1},
       {20, 0}, {21, 0}, {22, 1}, {23, 0}, {24, 0}, {25, 1}, {26, 0}, {27, 0}}
```

Figure A8.  Random draw used for simulations 7, 8 and 9.

In line 28 of Figure A8, twenty-seven random draws are taken from a Bernoulli distribution with parameter 0.2 which resulted in 6 ones. Note that 6/27 corresponds to 22.2 in Table 5. Here zeroes (ones) correspond to dingo-driven (tortoise-driven) vehicles. Thus, the output of line 30 indicates the first five drivers are dingo-like, the sixth driver is tortoise-like, drivers 7, 8 and 9 are dingo-like, and driver 10 is tortoise-like.

In Figure A9, red (black) points corresponds to dingo-driven (tortoise-driven) vehicles and we see Figures A7, A8 and A9 are consistent. Figure A9 shows the RGW-SIM display at the southern and northern ends of Telegraph Road and the time these snap shots were taken.

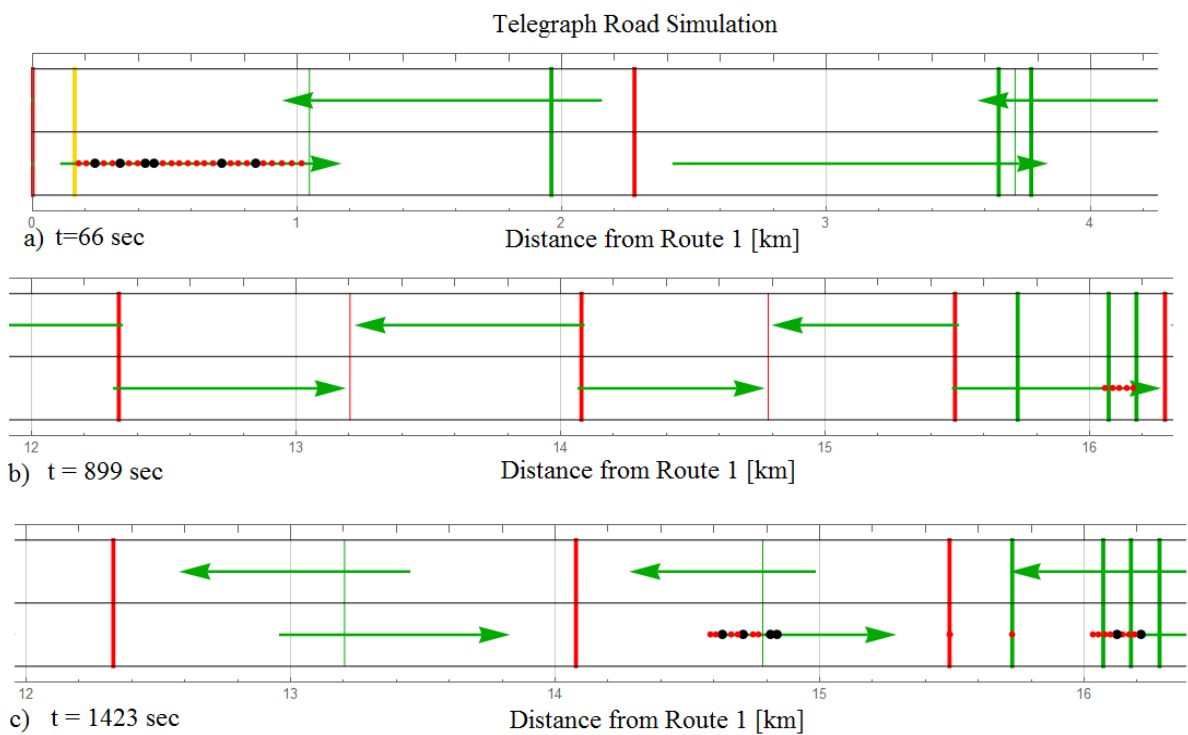

Figure A9. RGW-SIM display for simulation 9 at different instants of time. a) Time = 66 sec, b) Time = 899 sec, c) Time = 1423 sec.

In Figure A9 each red (black) dot corresponds to a vehicle driven by a dingo- (tortoise-) like driver. a) Shows this simulation starts out with 27 vehicles in the first green-wave. Each simulation in Table 5 with the exception of simulation 3 started out with 27 vehicles. b) Shows the first five dingoes at time equal to 899 sec shortly before they cross the Huntington-Telegraph intersection. c) Vehicles driven by tortoise-like drivers are about to cross the Huntington-Telegraph intersection significantly later than the dingo-like drivers illustrated in part b) of Figure A9.

*3. Method used to calculate Table 5 entries.* All entries in columns 6-12 were calculated from time-space diagrams similar to those exhibited in Figure A6. Figure A10 exhibits the time-space diagram DT1P3[t].

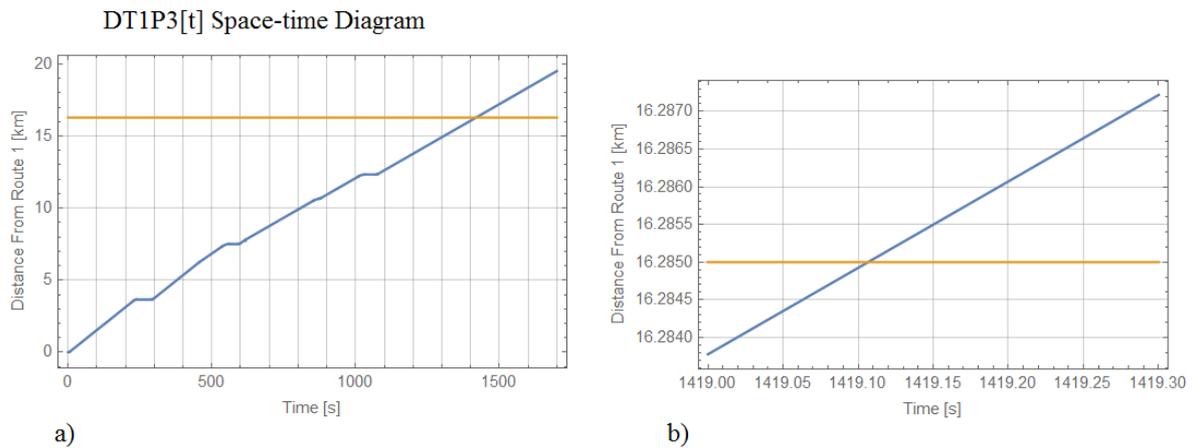

Figure A10. The DT1P3[t] time-space diagram. a) Showing entirety of time-space diagram on Telegraph Road between route 1 and Huntington Ave. b) Detailed view of time-space-diagram at Huntington Ave.

Part b) of Figure A10 shows that it takes 1419.1 seconds for vehicles described by DT1P3[t] to cross Huntington Ave while traveling from route 1 on Telegraph Road. Figure A7 shows that three vehicles driven by a dingo-like driver and one vehicle driven by a tortoise-like driver are described by DT1P3[t]. Mean travel times in Table 5 were weighted by the number of vehicles described by a particular time-space diagram. This is considerably lower than the 1389 sec value for the tortoise-dingo mixture because other time-space-time diagrams, not shown in Figure A7, raised the mean time to travel Telegraph Road.

Figure A10 shows that vehicles described by DT1P3[t] made three red light stops. This value contributed to the 4.4 average value for the mean number of traffic light stops indicated for the dingo-tortoise mixture shown in Table 5.

To calculate the mean time waiting at traffic lights given in column 8 of Table 5 it is necessary to first determine how much time each vehicle spent waiting at each traffic signal. Figure A11 shows how this was done for the first traffic light stop shown in Figure A10.

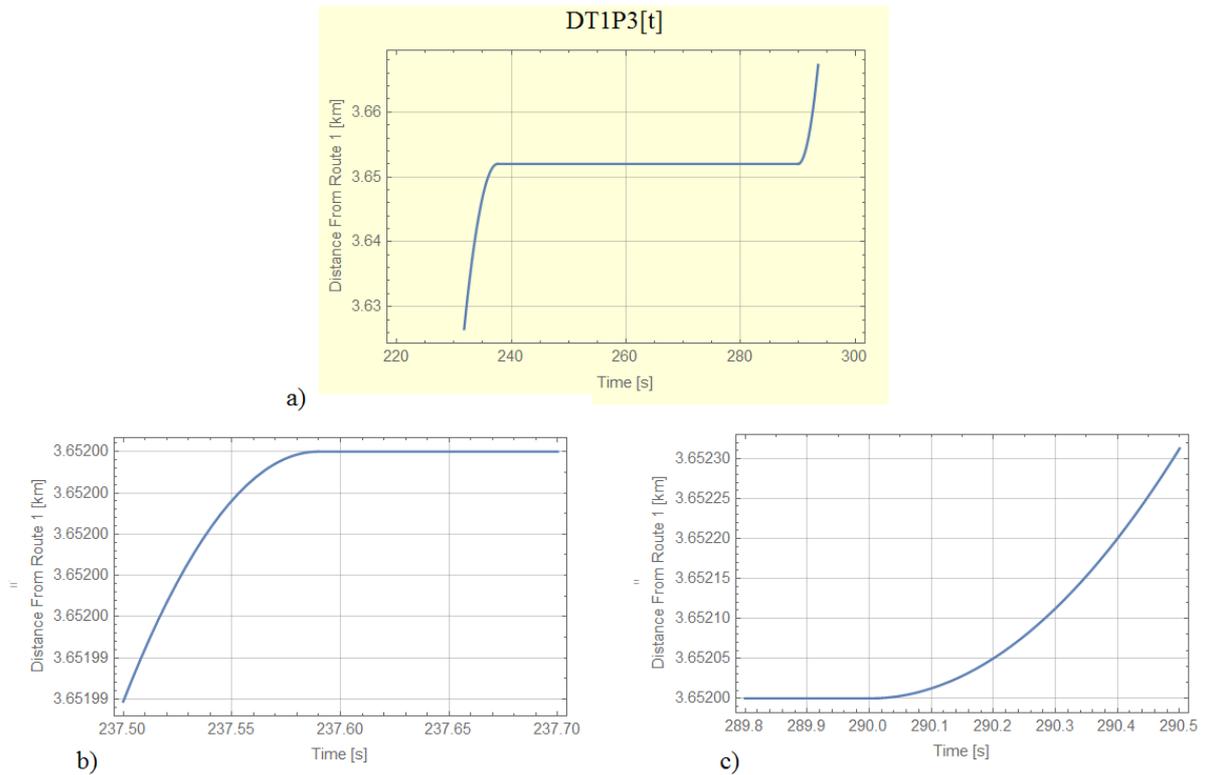

Figure A11. Method used to calculate vehicle mean time waiting at red light stops. a) Details of the first red light stop in Figure A10. b) Details of red-light stop. c) Details of red-light release.

Figure A11 shows vehicles described by DT1P3[t] were stopped at the red light for 52.4 seconds. This is much less than the value of 177 sec given in Table 5 because it is the time spent waiting for a single light and with the dingo-tortoise mixture, 4.4 is the average number of stops.